\documentclass[12pt]{article}
\usepackage{epsfig,cite}
\usepackage {amsmath}
\usepackage{amssymb}
\include{epsf}
\newcount\timecount
\newcount\hours \newcount\minutes  \newcount\temp \newcount\pmhours
\hours = \time
\divide\hours by 60
\temp = \hours
\multiply\temp by 60
\minutes = \time
\advance\minutes by -\temp
\def\hour{\the\hours}
\def\minute{\ifnum\minutes<10 0\the\minutes
            \else\the\minutes\fi}
\def\clock{
\ifnum\hours=0 12:\minute\ AM
\else\ifnum\hours<12 \hour:\minute\ AM
      \else\ifnum\hours=12 12:\minute\ PM
            \else\ifnum\hours>12
                 \pmhours=\hours
                 \advance\pmhours by -12
                 \the\pmhours:\minute\ PM
                 \fi
            \fi
      \fi
\fi
}

\def\monthname{\relax\ifcase\month 0/\or January\or February\or
   March\or April\or May\or June\or July\or August\or September\or
   October\or November\or December\else\number\month/\fi}

\def\bold#1{\setbox0=\hbox{$#1$}%
     \kern-.025em\copy0\kern-\wd0
     \kern.05em\copy0\kern-\wd0
     \kern-.025em\raise.0433em\box0 }

\def\beq{\begin{equation}}
\def\eeq{\end{equation}}

\def\ga{\mathrel{\raise.3ex\hbox{$>$\kern-.75em\lower1ex\hbox{$\sim$}}}}
\def\la{\mathrel{\raise.3ex\hbox{$<$\kern-.75em\lower1ex\hbox{$\sim$}}}}
\def\gev{{\rm \, Ge\kern-0.125em V}}
\def\tev{{\rm \, Te\kern-0.125em V}}
\def\gyr{{\rm \, G\kern-0.125em yr}}

\def\tb{\tan \beta}

\def\gappeq{\mathrel{\rlap {\raise.5ex\hbox{$>$}}
{\lower.5ex\hbox{$\sim$}}}}
\def\lappeq{\mathrel{\rlap{\raise.5ex\hbox{$<$}}
{\lower.5ex\hbox{$\sim$}}}}
\def\Toprel#1\over#2{\mathrel{\mathop{#2}\limits^{#1}}}

\def\m12{m_{1\!/2}}

\def\bea{\begin{eqnarray}}
\def\eea{\end{eqnarray}}

\begin{document}
\begin{titlepage}
\pagestyle{empty}
\baselineskip=21pt
\rightline{KCL-PH-TH/2012-08, LCTS/2012-04, CERN-PH-TH/2012-045}
\rightline{UMN--TH--3034/12, FTPI--MINN--12/07}
\vskip 0.2in
\begin{center}
{\large{\bf Revisiting the Higgs Mass and Dark Matter in the CMSSM}}
\end{center}
\begin{center}
\vskip 0.3in
{\bf John~Ellis}$^1$
and {\bf Keith~A.~Olive}$^2$
\vskip 0.3in
{\small {\it
$^1${Theoretical Particle Physics and Cosmology Group, Department of
  Physics, King's~College~London, London WC2R 2LS, United Kingdom;\\
Theory Division, CERN, CH-1211 Geneva 23,
  Switzerland}\\
$^2${William I. Fine Theoretical Physics Institute, School of Physics and Astronomy,\\
University of Minnesota, Minneapolis, MN 55455, USA}\\
}}

\vskip 0.25in
{\bf Abstract}
\end{center}
\baselineskip=18pt \noindent

Taking into account the available accelerator and astrophysical constraints,
the mass of the lightest neutral Higgs boson $h$ in the
minimal supersymmetric extension of the Standard Model with universal soft
supersymmetry-breaking masses (CMSSM) has been estimated to lie between 114 and $\sim 130$~GeV.
Recent data from ATLAS and CMS hint that $m_h \sim 125$~GeV, 
though $m_h \sim 119$~GeV may still be a possibility. Here we study the consequences
for the parameters of the CMSSM and direct dark matter detection if the Higgs hint is confirmed,
focusing on the strips in the $(m_{1/2}, m_0)$ planes for different $\tb$ and $A_0$ 
where the relic density of the lightest neutralino $\chi$ falls
within the range of the cosmological cold dark matter
density allowed by WMAP and other experiments. 
We find that if $m_h \sim 125$~GeV focus-point
strips would be disfavoured, 
as would the low-$\tb$ ${\tilde \tau}-\chi$ and ${\tilde t}_1 -\chi$ coannihilation strips,
whereas the ${\tilde \tau}-\chi$ coannihilation strip at large $\tb$ and $A_0 > 0$ 
would be favoured, together with its extension to a funnel where rapid annihilation via
direct-channel $H/A$ poles dominates. On the other hand,
if $m_h \sim 119$~GeV more options would be open. We give parametrizations of WMAP
strips with large $\tb$ and fixed $A_0/m_0 > 0$ that include 
portions compatible with $m_h = 125$~GeV, and present predictions for spin-independent
elastic dark matter scattering along these strips. These are generally low for
models compatible with $m_h = 125$~GeV, whereas the XENON100 experiment already
excludes some portions of strips where $m_h$ is smaller.

\vfill
\leftline{February 2012}
\end{titlepage}
\baselineskip=18pt

\section{Introduction}

Since supersymmetry relates the Higgs self-coupling to electroweak gauge couplings,
it is a characteristic prediction of the minimal supersymmetric extension of
the Standard Model (MSSM) that the lightest neutral Higgs boson $h$
should be relatively light. This prediction is in agreement with the indirect indications from 
precision electroweak data, which favour $m_h \sim 100$~GeV \cite{Gfitter}. 
Indeed, at the tree level $m_h$ would be $< m_Z$, but radiative corrections
due principally to the top squarks may increase $m_h$ to $\sim 130$~GeV within the
MSSM \cite{mh}. The latest LHC searches for a Standard Model-like Higgs boson exclude the
mass range $m_h \in (127, 600)$~GeV at the 95\% CL, but leave open the range
$m_h \in (115.5, 127)$~GeV, which is consistent with the MSSM prediction \cite{ATLAS+CMS}.
Moreover, within this range ATLAS sees an excess of events with $m_h \sim 126$~GeV \cite{Dec13a},
and CMS sees a broader excess extending over the range $m_h \in (119, 125)$~GeV \cite{Dec13c}.
The observations of these excesses are more significant within the MSSM than in the
general Standard Model context, since the look-elsewhere effect is diminished for the
restricted Higgs mass range predicted previously within the MSSM.

Identifying the lightest neutralino $\chi$ as the dominant component of dark matter
provides an important constraint on the MSSM parameter space \cite{EHNOS} that can be
quantified within any specific framework for supersymmetry breaking. Here we
assume the CMSSM \cite{funnel,cmssm,efgosi,cmssm2,eoss,cmssmwmap,mc1-3}, 
in which the soft supersymmetry-breaking parameters 
$m_{1/2}, m_0$ and $A_0$ are constrained to be universal at the GUT scale.
In the CMSSM, there are narrow strips of parameter space where the relic $\chi$
density falls within the narrow range indicated by WMAP and other astrophysical
and cosmological measurements \cite{wmap}, which we take to be $\Omega_\chi h^2 = 0.112 \pm 0.12$
corresponding to a conservative $2-\sigma$ range. These strips include one where coannihilation
between $\chi$, the lighter stau ${\tilde \tau_1}$ and other sleptons brings the
cold dark matter density into the WMAP range \cite{efo}, which at large $\tan \beta$ extends 
into a funnel where annihilations via direct-channel $H/A$ resonances are
dominant \cite{funnel,efgosi}, and a focus-point strip at large $m_0$ where annihilation is enhanced
by a significant Higgsino component in the composition of $\chi$ \cite{fp}. When $A_0$ is
large, there may also be a strip where $\chi$ coannihilation with the lighter stop
${\tilde t_1}$ is important \cite{stopco}.

These strips are useful for benchmarking searches for supersymmetry
at colliders~\cite{bench} and searches for astrophysical dark matter~\cite{EFFMO}, e.g., via
direct searches for elastic scattering~\cite{EOS} or via searches for energetic neutrinos
produced by dark matter annihilations in the core of the Sun or Earth~\cite{EOSSnu}, or via
searches for energetic photons from dark matter annihilations near the centre of the
Galaxy~\cite{EOSp} or elsewhere. It is therefore important that the benchmark strips
should be updated to take the latest accelerator and other constraints into account~\cite{recentbench}.
This is the main purpose of this paper, with particular focus on highlighting (portions of)
strips that are compatible with a hypothetical measurement of $m_h$, and their implications
for dark matter detection.

It is a general feature of relic density calculations that they yield an upper limit
on the magnitudes of $m_{1/2}$ and $m_0$, and hence on $m_h$. This
connection was explored in~\cite{ENOS}, with the result that an upper bound
$m_h \sim 127$~GeV was found within the CMSSM, after the experimental,
phenomenological, astrophysical and cosmological constraints then
available were taken into account. This result was confirmed in a recent
global frequentist analysis of constraints on the CMSSM \cite{mc1-3,mc5-6,MC7,mc7.5}. 
It is encouraging
that the excesses found by ATLAS and CMS fall within the range allowed by
these analyses.

Within a specific framework such as the CMSSM, a measurement of $m_h$
would impose a complementary constraint on the strips of parameter space
allowed by the dark matter density, although with some uncertainty due to
the error $\sim \pm 1.5$~GeV associated with the theoretical calculation of $m_h$
for any given value of the CMSSM parameters~\cite{FeynHiggs}. The impact of this constraint on
the CMSSM was explored recently \cite{mc7.5,post-mh}, for the values $m_h \sim 125$ and 119~GeV 
suggested by the recent ATLAS and CMS results. It was shown, in particular, 
that relatively large values of $m_{1/2}, m_0, A_0$ and $\tan \beta$ would be
favoured if $m_h \sim 125$~GeV.

In this paper we explore in more detail the potential implications of an LHC
measurement of $m_h \sim 125$ or 119~GeV for the WMAP-compatible strips
of the CMSSM, concentrating on the case $\mu > 0$. We find that a measurement of $m_h \sim 125$~GeV would
favour the ${\tilde \tau}_1 - \chi$ coannihilation strip and its extension to the
rapid $H/A$ annihilation funnel at large $\tan \beta$ with $A_0 > 0$, and disfavour the
low-$\tb$ ${\tilde \tau}_1 - \chi$ coannihilation strip, as well as the ${\tilde t_1}-\chi$
coannihilation and focus-point strips (except if $m_0 > 5000$~GeV). On the other hand, a
measurement of $m_h \sim 119$~GeV would keep these options open. We also
discuss the interplay within the CMSSM between $m_h$ and the
elastic dark matter scattering cross section. We find that, whereas some low-$m_h$
models are already excluded by the XENON100 experiment~\cite{XENON100}, models with  $m_h \sim 125$~GeV
typically predict cross sections well below the present experimental sensitivity.

\section{Summary of Results from Scans of the CMSSM Parameter Space}

\begin{figure}
\vskip 0.5in
\vspace*{-0.75in}
\begin{minipage}{8in}
\epsfig{file=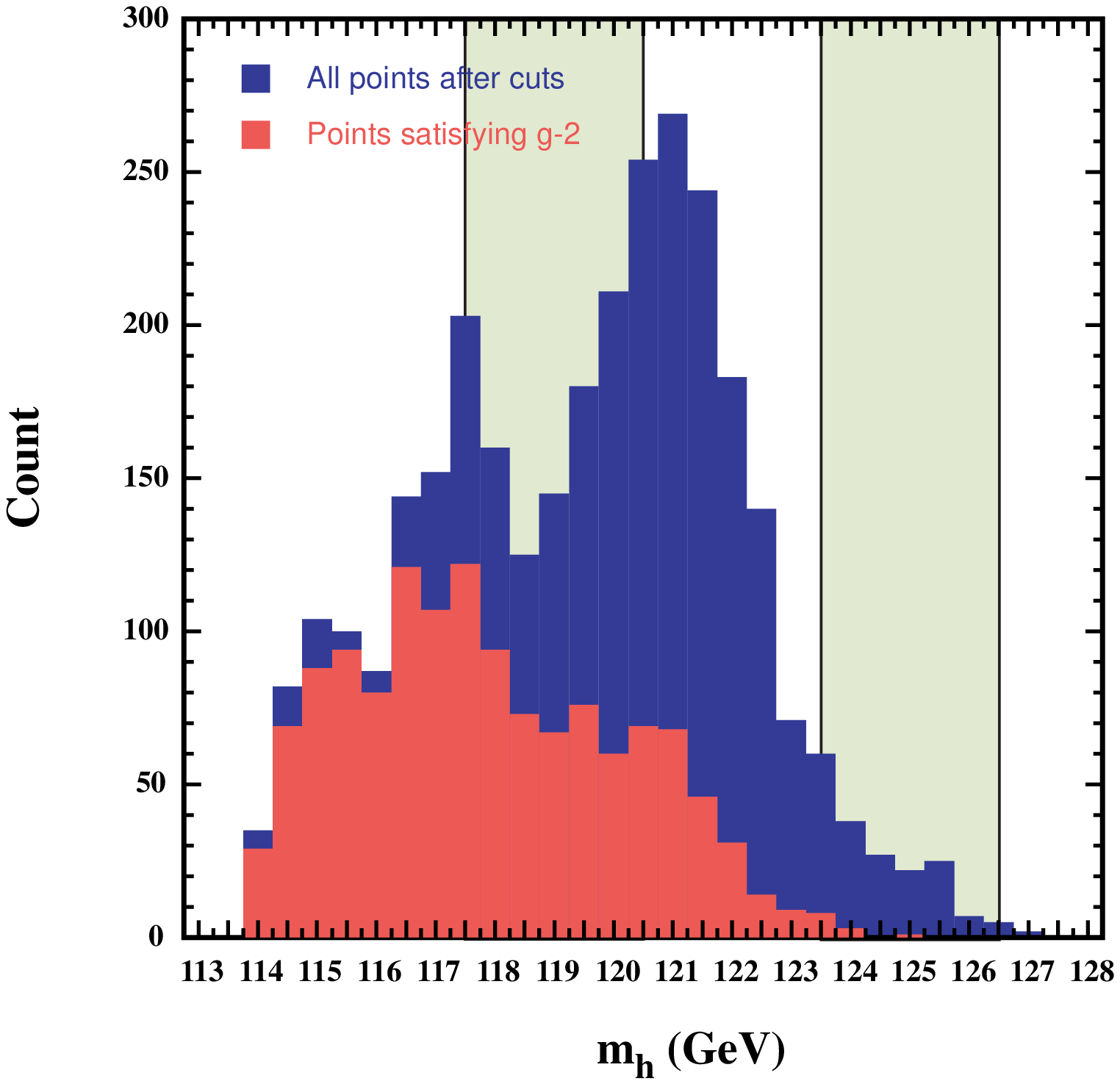,height=3.1in}
\epsfig{file=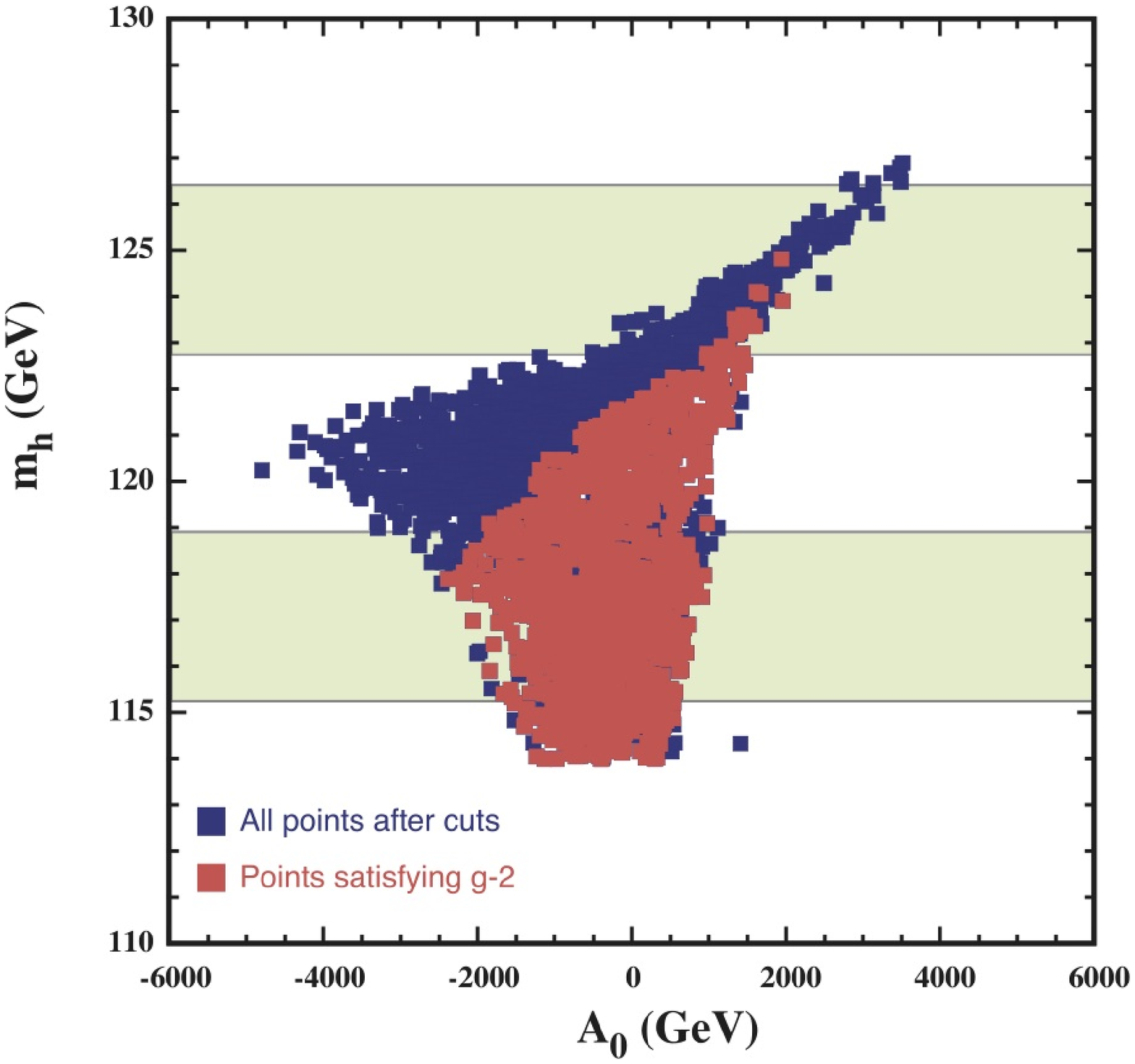,height=3.3in}
\hfill
\end{minipage}
\begin{minipage}{8in}
\epsfig{file=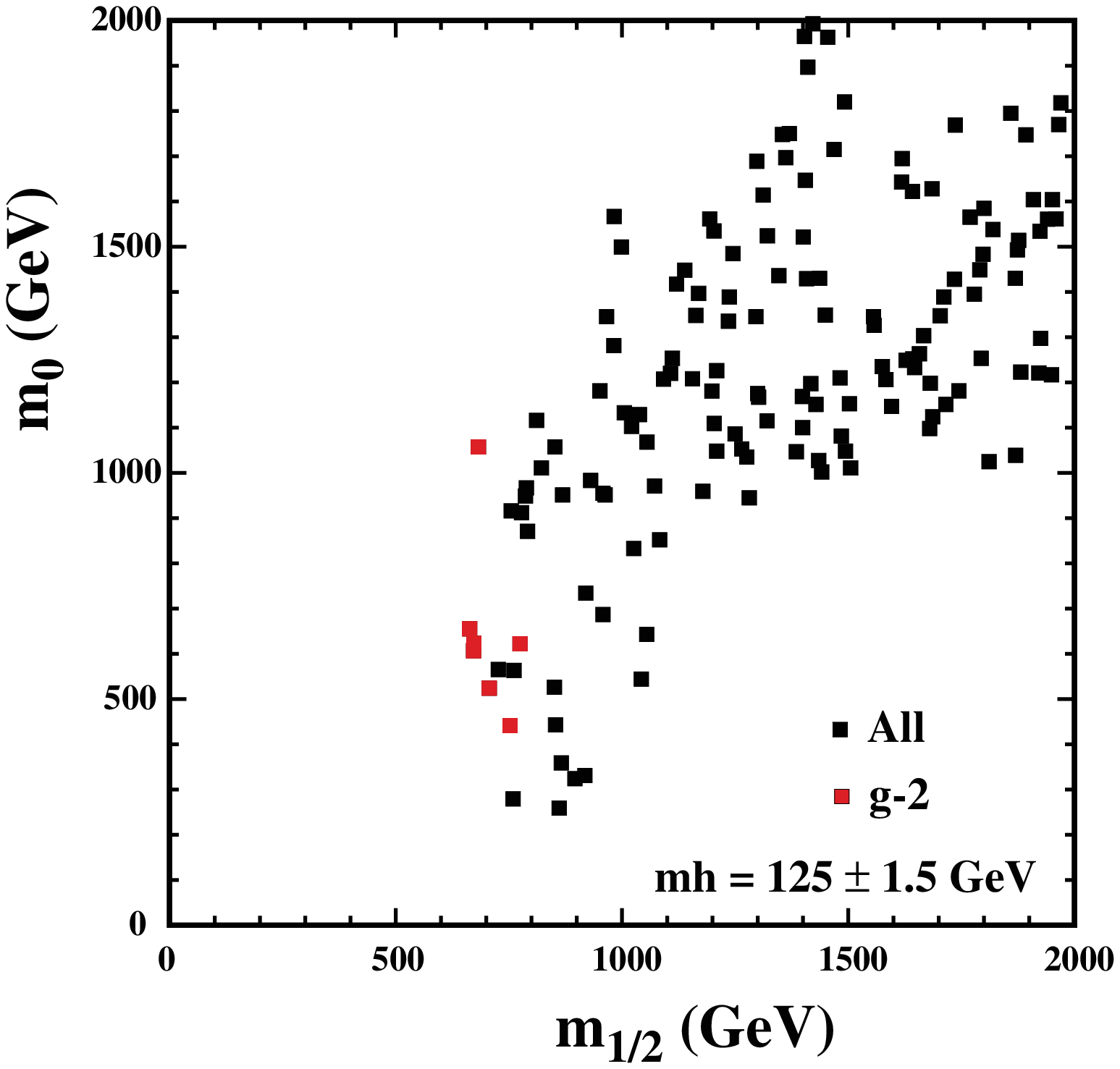,height=3.1in}
\epsfig{file=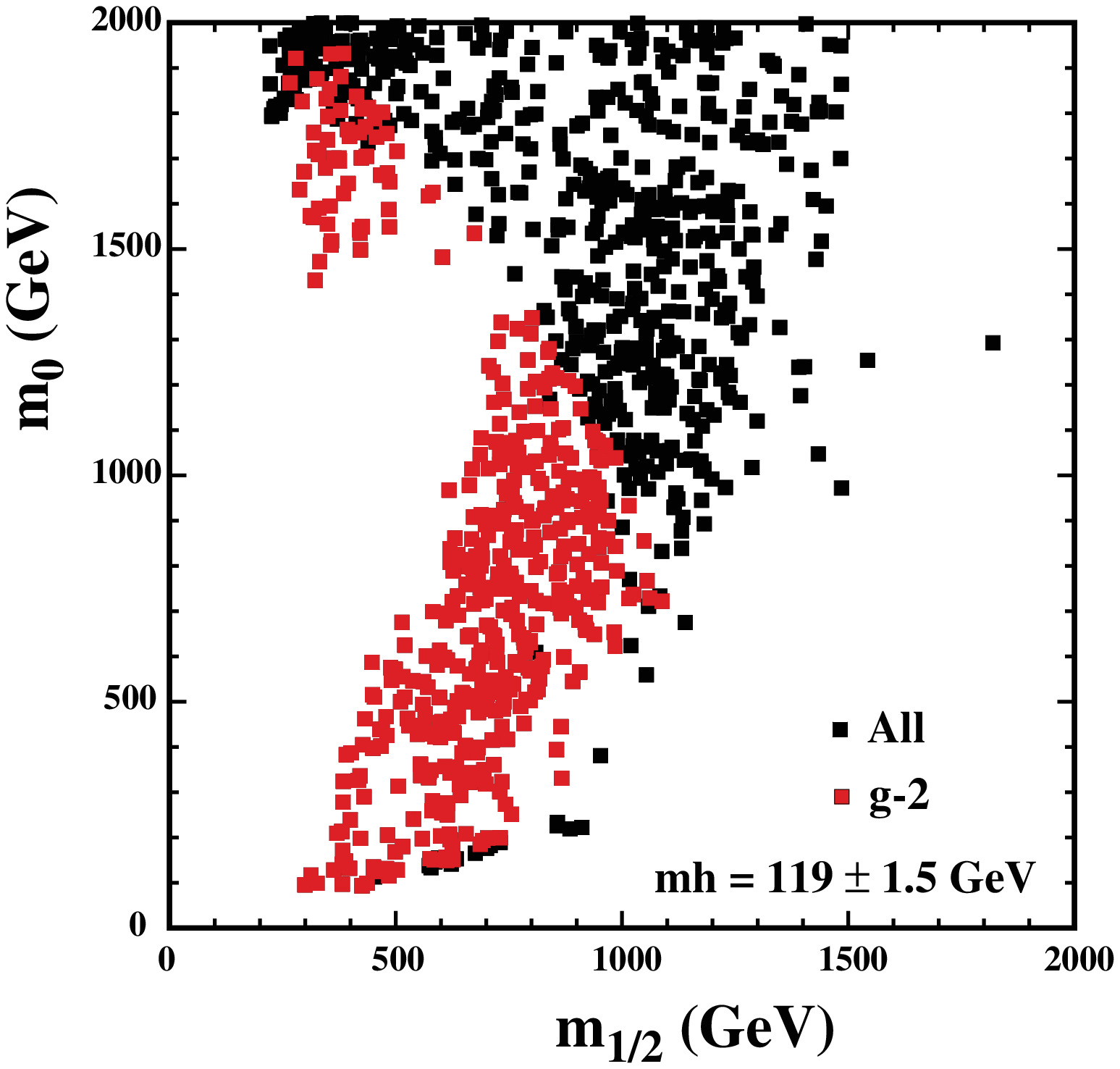,height=3.1in}
\hfill
\end{minipage}
\caption{
{\it
Upper panels: Histogram of values of $m_h$ found in a pre-LHC scan of the CMSSM
parameter space~\protect\cite{ENOS} (left), 
and displaying the corresponding values of $A_0$ (right). In both panels we have added
green bands corresponding to the ranges $119 \pm 1.5$~GeV and $125 \pm 1.5$~GeV 
hinted by the LHC~\protect\cite{Dec13a,Dec13c}.
Lower panels: The distributions of the points from~\protect\cite{ENOS} in the $(m_{1/2}, m_0)$ plane
of the CMSSM for which {\tt FeynHiggs} yields
$m_h = 125 \pm 1.5$~GeV (left) and $119 \pm 1.5$~GeV (right).
These results were obtained assuming $m_t = 174.3$~GeV.
Points favoured by the $g_\mu - 2$ constraint are highlighted in red in all four panels.
}} 
\label{fig:mh}
\end{figure}

As mentioned in the Introduction, in 2005 a scan of the CMSSM
parameter space  was made over the ranges 100~GeV $< m_{1/2} < 2$~TeV, $m_0 < 2$~TeV, 
$|A_0/m_{1/2}| < 3$, $2 < \tan \beta < 58$ and $\mu > 0$, mostly with $m_t = 174.3$~GeV
though other values of $m_t$ were also considered in less detail. The principal result
of this scan was a histogram of $m_h$ shown in Fig.~1 of~\cite{ENOS},
which displayed a range extending up to $\sim 127$~GeV. This is
reproduced in the upper left panel of Fig.~\ref{fig:mh},
with vertical green bands added to indicate the ranges $m_h = 119 \pm 1.5$~GeV and $125 \pm 1.5$~GeV
hinted by LHC data~\cite{Dec13a,Dec13c}. It is encouraging that the range found in~\cite{ENOS}
includes the value $m_h \sim 125$~GeV currently preferred by ATLAS and CMS. However,
it is equally clear that this value is far from the mode of the histogram. The lower left panel
of Fig.~\ref{fig:mh} displays the (relatively few) points compatible with $m_h = 125$~GeV as calculated using
{\tt FeynHiggs}~\cite{FeynHiggs} within the theoretical error of $\pm 1.5$~GeV, 
highlighting (in red) the (very few) points favoured by $g_\mu - 2$~\cite{newBNL} at the $2 - \sigma$ level. We see that most of the
points compatible with $m_h = 125$~GeV have $m_{1/2}$ and $m_0$ both $> 1$~TeV, whereas the $g_\mu - 2$-compatible points
are concentrated at small values of $m_{1/2}$ and $m_0$. We do not reproduce here the histogram
of values of $\tan \beta$ shown in Fig.~2 of~\cite{ENOS}, but recall that it was concentrated at $\tan \beta > 50$,
with a tail extending down to $\tan \beta \sim 10$. The upper right panel of Fig.~\ref{fig:mh} shows 
that the CMSSM points compatible with $m_h \sim 125$~GeV are concentrated at large values of $A_0 > 0$.

On the other hand, it is clear from the upper left panel of Fig.~\ref{fig:mh} that the 2005 scan
found more points with $m_h \sim 119$~GeV (though this was also not a mode of the $m_h$
histogram!) and we see from the lower right panel of Fig~\ref{fig:mh} that many of these points
had $m_{1/2}$ and/or $m_0 < 1$~TeV. As seen in the upper right panel of Fig.~\ref{fig:mh},
points with $m_h \sim 119$~GeV exhibit no preference for either sign of $A_0$.

Recently, as members of the {\tt MasterCode} collaboration, we have
participated in a frequentist analysis of the relative likelihoods of different
points in the CMSSM parameter space incorporating the constraints from
LHC missing-energy searches~\cite{LHCMET} on supersymmetric particles with $\sim 1$/fb of data~\cite{MC7}.
This analysis favoured $m_h \sim 119$~GeV, with a likelihood price $\Delta \chi^2 \sim 2$
for a hypothetical measurement $m_h \sim 125$~GeV. As shown in~\cite{mc7.5},
such a value of $m_h$ would indicate within the CMSSM a preference for relatively 
large values of $m_{1/2}, m_0$ and $\tan \beta$, confirming the results of~\cite{ENOS}.
However, we also note that smaller values of $\tan \beta \sim 10$ are still allowed at the
68\% CL even if $m_h \sim 125$~GeV, in association with $m_{1/2} \sim 700$~GeV.
The analysis of this paper complements that of~\cite{mc7.5}, by providing more insight
into the interplay of the principal constraints and the resulting predictions for direct
dark matter detection.

\section{Dark Matter Strips}

As was reviewed in the Introduction, the requirement that the relic neutralino
density falls within the range allowed by WMAP and other observations implies that,
the allowed values of $m_{1/2}$ and $m_0$ lie along narrow strips in generic 
$(m_{1/2}, m_0)$ planes for fixed values of $A_0$ and $\tan \beta$ \cite{eoss}. Along these
strips, the dominant mechanism fixing the relic density may be coannihilation with some
near-degenerate sparticle species, such as the ${\tilde \tau}_1$
or ${\tilde t}_1$, or $\chi - \chi$ annihilations facilitated by 
direct-channel heavy Higgs $H/A$ poles (in rapid-annihilation funnels) or by enhanced Higgsino components
(along focus-point strips). As was also mentioned in the Introduction,
points along these WMAP-compatible strips are often used as benchmarks~\cite{bench} for
dark matter searches~\cite{EFFMO}, e.g., via scattering~\cite{EOS} or annihilations into neutrinos~\cite{EOSSnu}
or photons~\cite{EOSp}. These benchmarks strips require updating in light of the strengthening LHC
constraints on supersymmetry~\cite{recentbench} and the hypothetical Higgs mass measurement~\cite{Dec13a,Dec13c}.

\begin{figure}
\vskip 0.5in
\vspace*{-0.75in}
\begin{minipage}{8in}
\epsfig{file=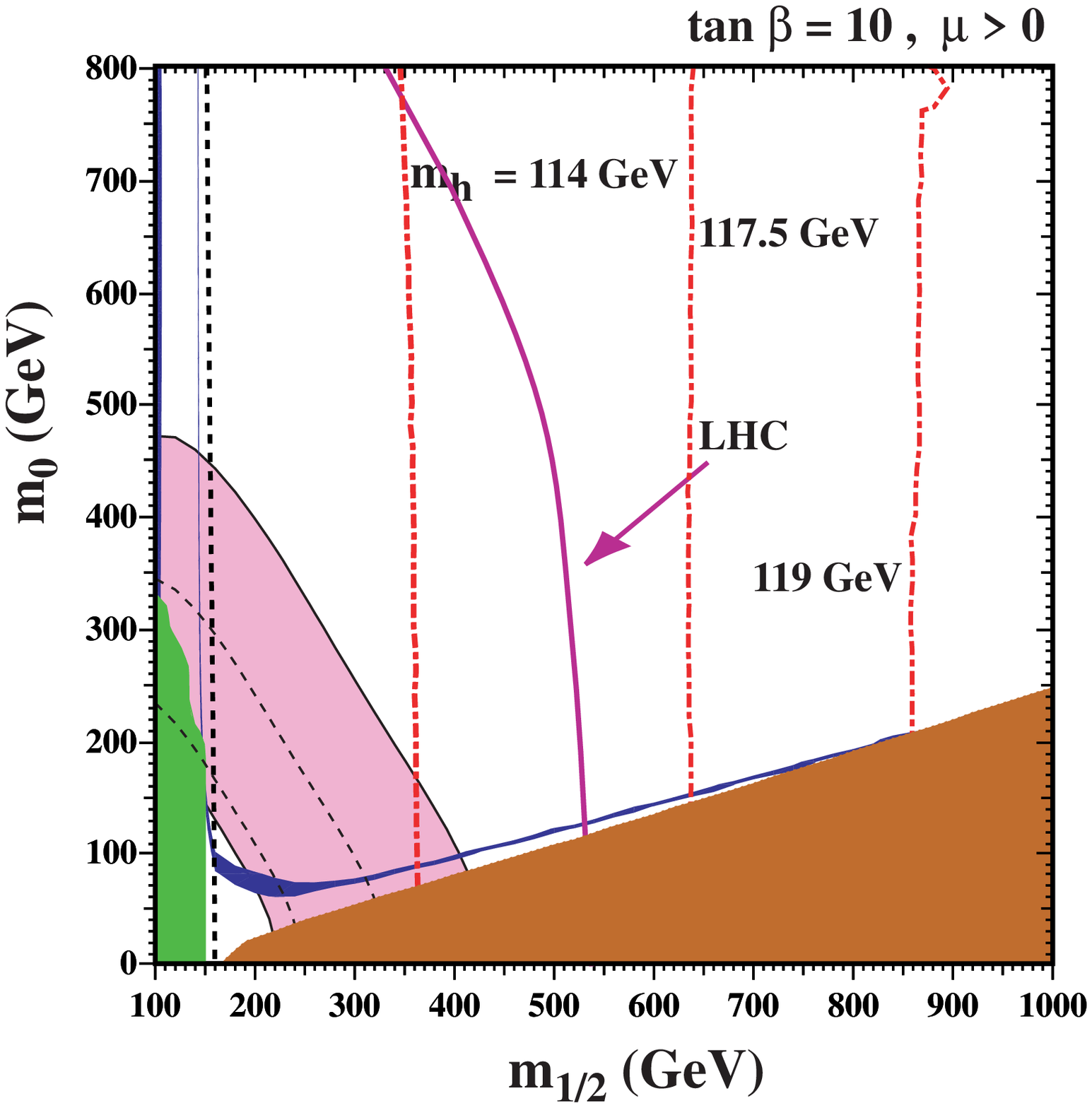,height=3.3in}
\hspace*{-0.17in}
\epsfig{file=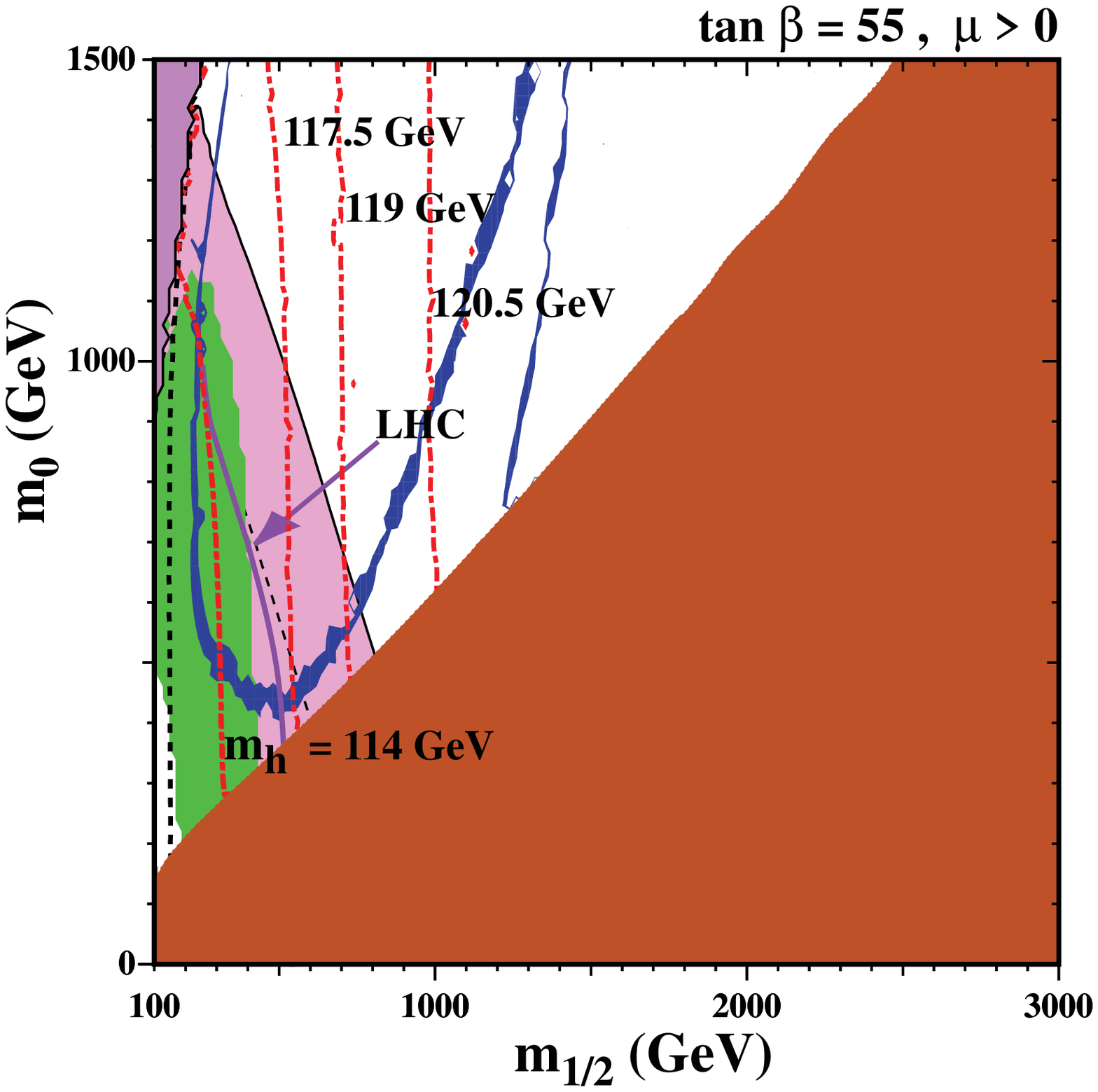,height=3.3in}
\hfill
\end{minipage}\caption{\it
The $(m_{1/2}, m_0)$ planes for $\mu > 0$, $A_0 = 0$ and
$\tan \beta = 10$ (left) and 55 (right), as 
calculated for $m_t = 173.2$~GeV using the latest version of the {\tt 
SSARD} code~\protect\cite{SSARD}. The WMAP strips where $\Omega_\chi h^2 = 0.112 \pm 0.012$
are shaded dark blue: note the narrow coannihilation strip in the left panel 
and the coannihilation strip, rapid-annihilation funnel and focus-point strip
in the right panel. The other shadings and colours of the contours
are described in the text.} 
\label{fig:strips} 
\end{figure}

Fig.~\ref{fig:strips} displays the latest incarnations of the $(m_{1/2}, m_0)$
planes for $\mu > 0$, $A_0 = 0$ and $\tan \beta = 10$ (left) and 55 (right)~\footnote{As already mentioned,
we focus here on $\mu > 0$. This assumption was motivated in the past by indications from $g_\mu - 2$ and the desire
to avoid strong constraints from $b \to s \gamma$ \cite{bsg}, but should perhaps be reviewed now in light of the growing
tension between LHC missing-energy constraints~\cite{LHCMET} and $g_\mu - 2$~\cite{newBNL}.}. Here and in subsequent
figures, the regions forbidden because the LSP is charged are 
shaded brown, the regions where there is no consistent electroweak vacuum are
shaded (darker) pink, the regions excluded by $b \to s \gamma$ \cite{bsg} are shaded
green, the regions favoured by $g_\mu - 2$~\cite{newBNL} at the $\pm 2 - \sigma$ level are shaded
(paler) pink, and the WMAP-compatible dark matter strips are shaded dark blue.
The black dashed line is the $m_{\chi^\pm_1} = 104$~GeV contour,
the solid purple lines outline the 95\% CL constraints on $(m_{1/2}, m_0)$ in the CMSSM
imposed by missing-energy searches at the LHC~\cite{LHCMET},
and contours of $m_h$ as calculated using {\tt FeynHiggs} \cite{FeynHiggs} are shown as red dash-dotted lines. Here, and in all subsequent analyses, a top quark mass of 173.2 GeV was used \cite{mt1731}. 
We see that $m_h = 119 \pm 1.5$~GeV is compatible with the dark matter constraint only
for $m_{1/2} > 640$~GeV if $\tan \beta = 10$, and for $m_{1/2} > 560$~GeV if $\tan \beta = 55$.
These regions are in the upper portions of the ${\tilde \tau}_1 - \chi$ coannihilation strips, 
which extend to $m_{1/2} \sim 900$~GeV.
The contours $m_h = 125 \pm 1.5$~GeV are nowhere to be seen as, e.g., nominal {\tt FeynHiggs} values
of $m_h$ do not exceed $\sim 120$~GeV for $m_0 < 3000$~GeV along the focus-point strip
for $\tb = 55$ and $A_0 = 0$.

\begin{figure}[htb!]
\begin{minipage}{8in}
\epsfig{file=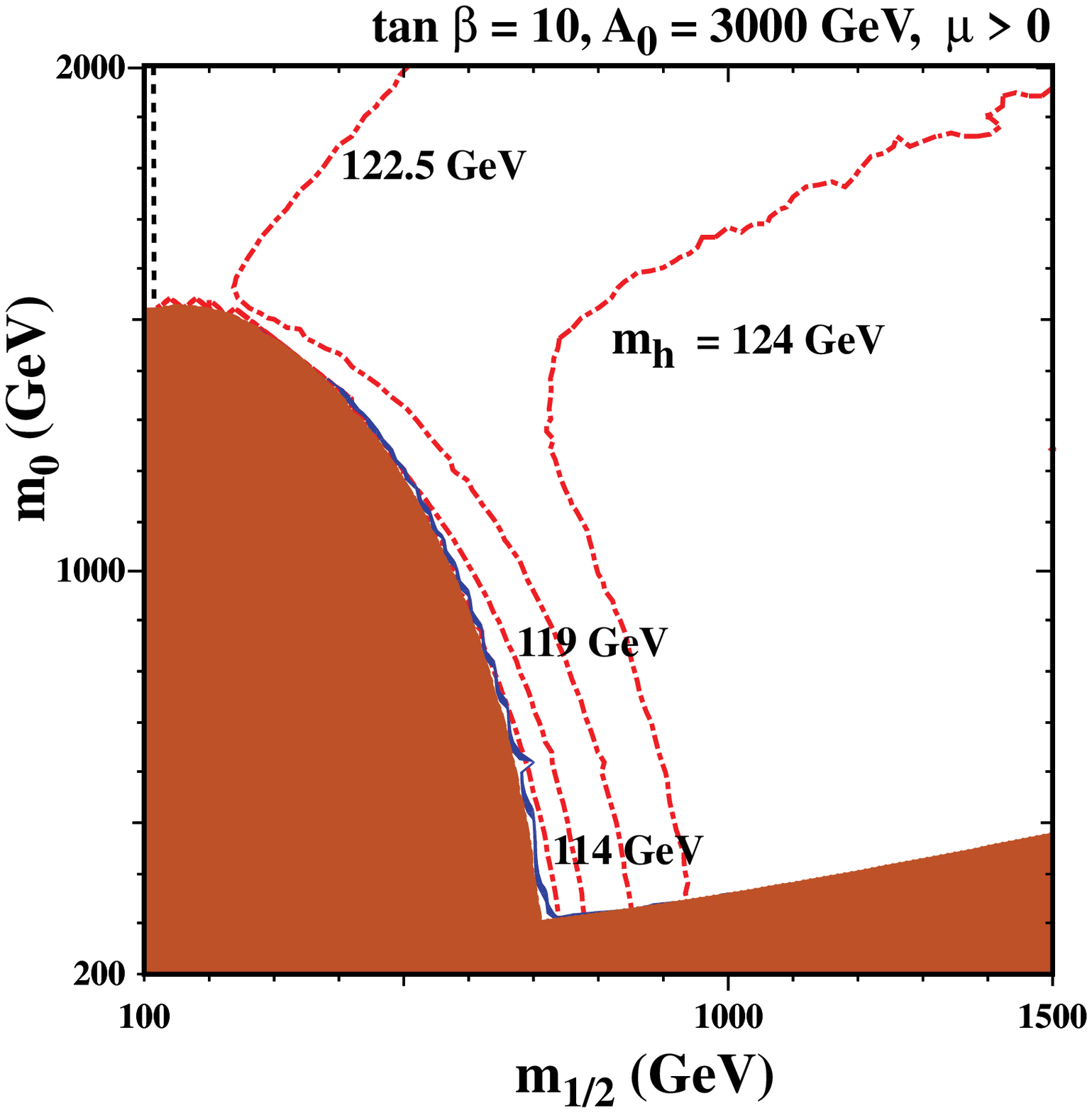,height=3.3in}
\hspace*{-0.17in}
\epsfig{file=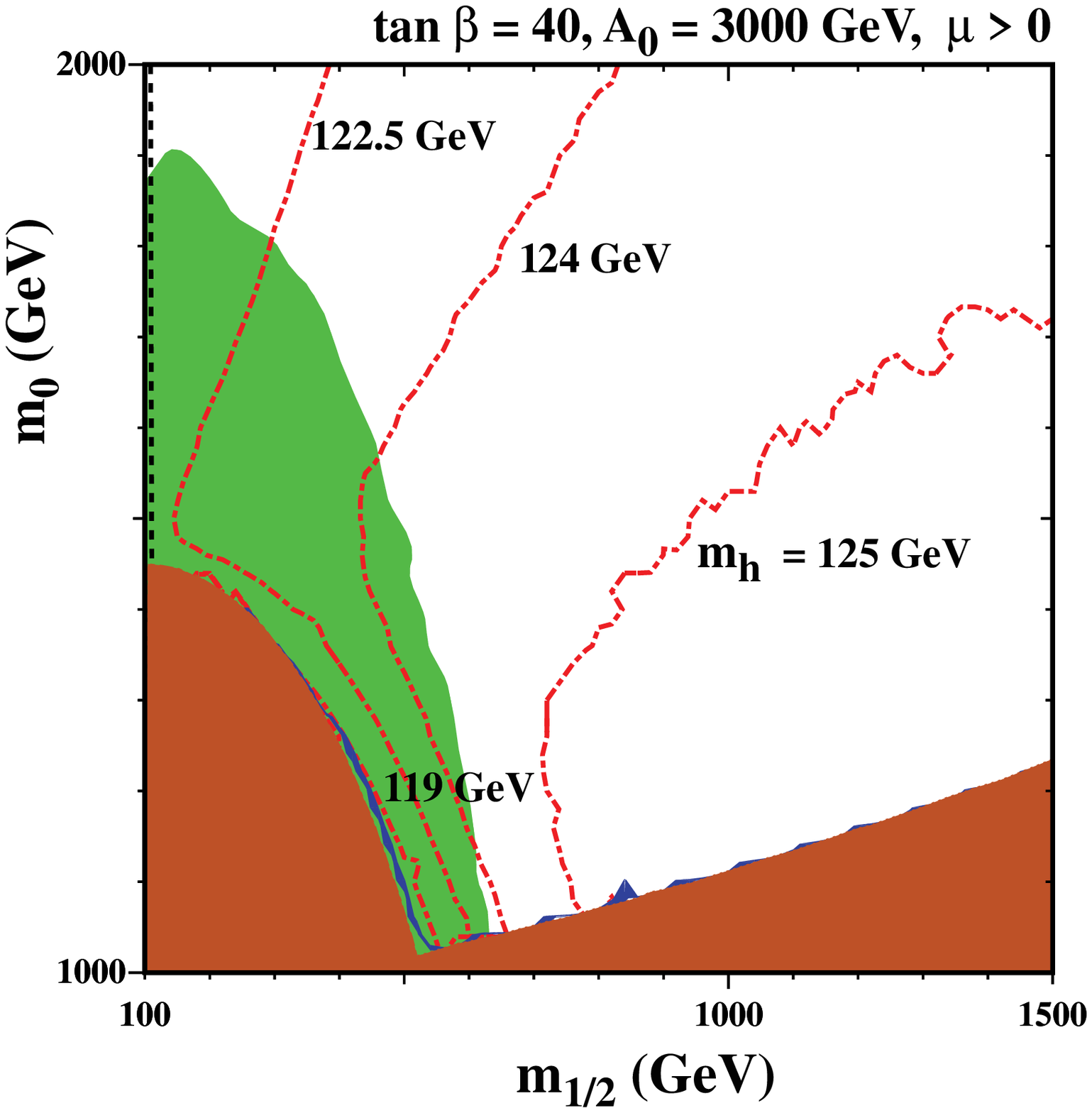,height=3.3in}
\hfill
\end{minipage}
\begin{minipage}{8in}
\epsfig{file=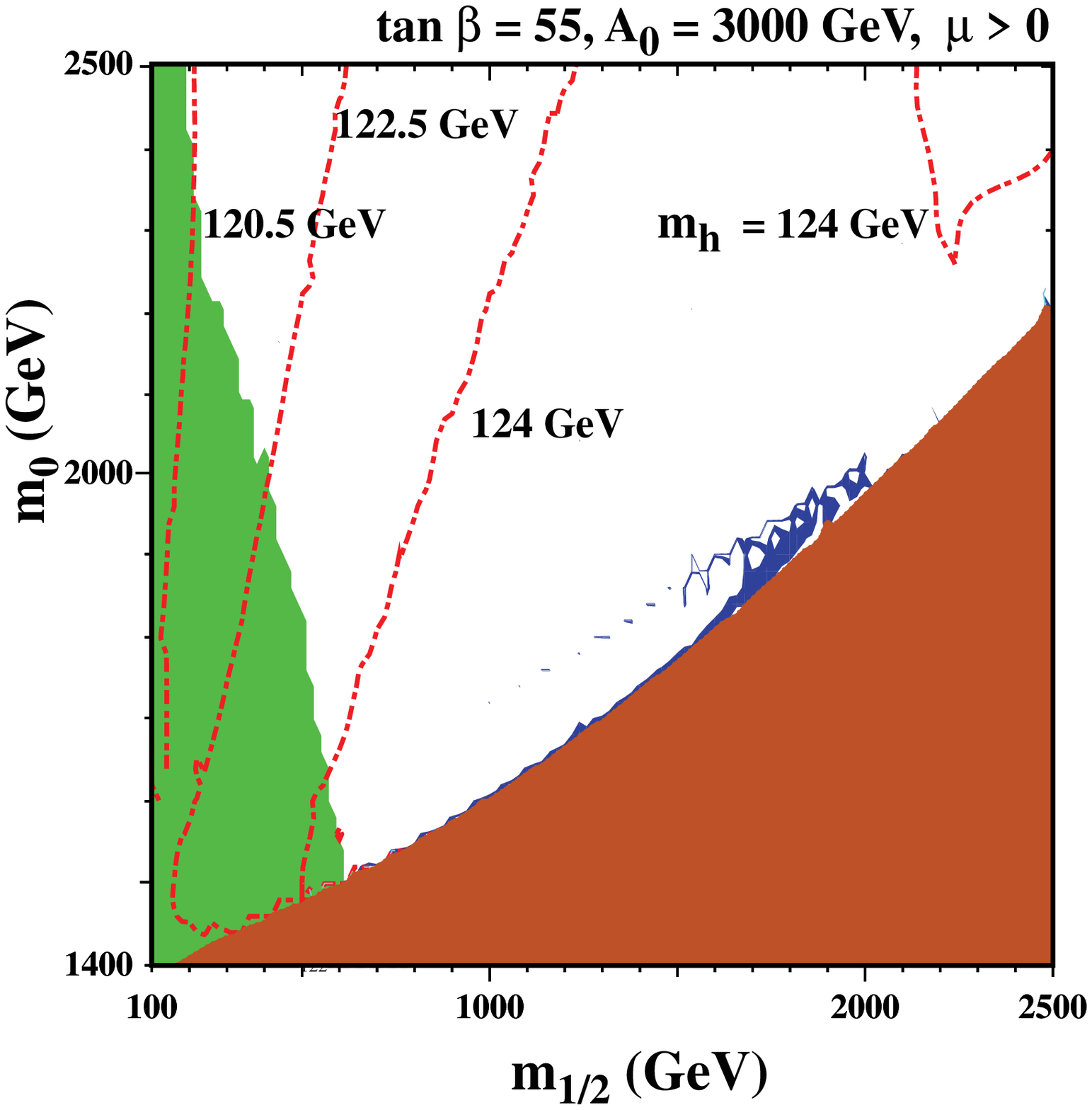,height=3.3in}
\hspace*{-0.17in}
\epsfig{file=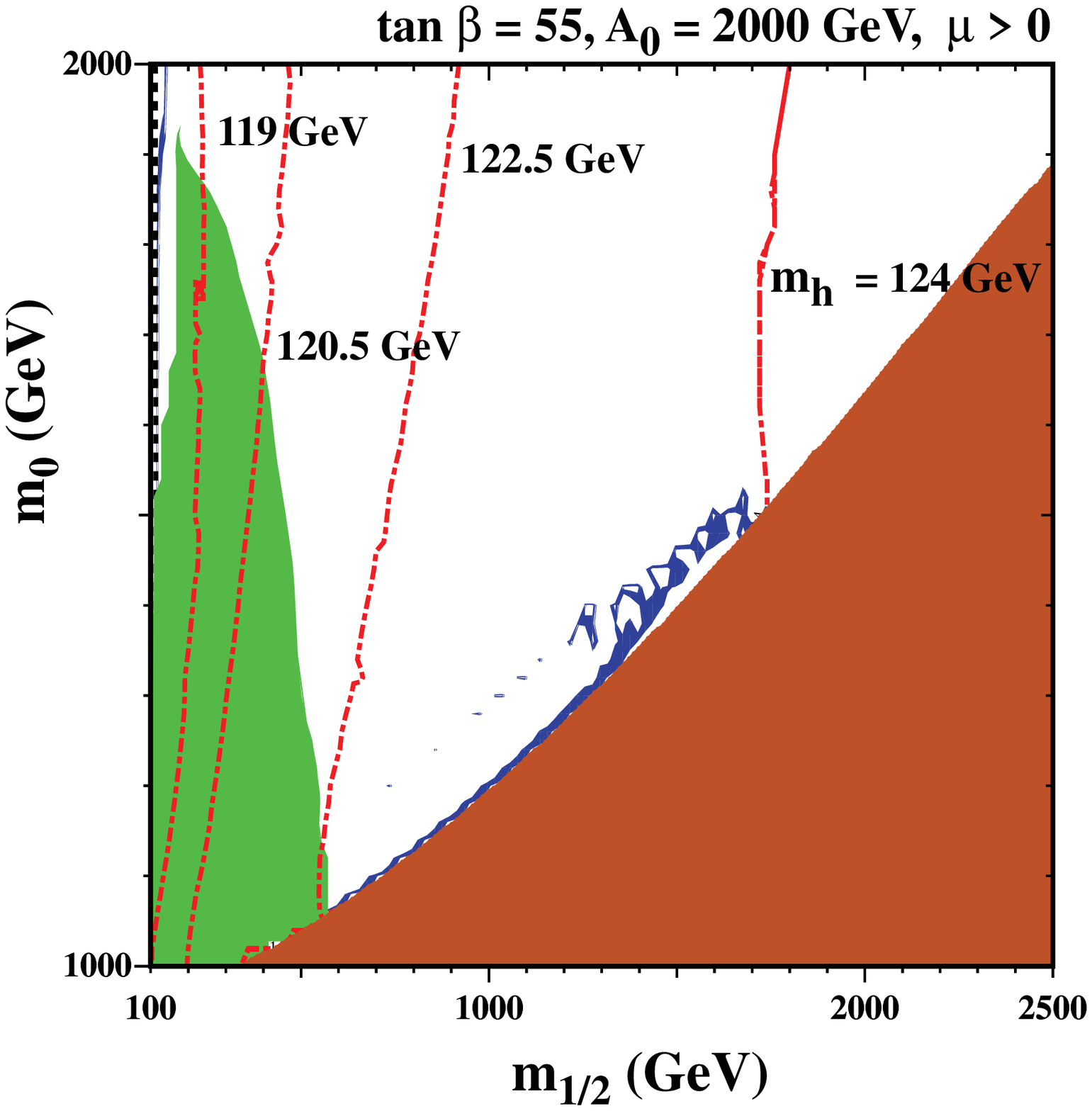,height=3.3in}
\hfill
\end{minipage}\caption{
{\it
As Fig.~\protect\ref{fig:strips}, with $\tan \beta = 10$ in the upper left panel,
$\tan \beta = 40$ in the upper right panel, and $\tan \beta = 55$ in the lower panels.
We choose $A_0 = 3000$~GeV, except in the lower right panel where $A_0 = 2000$~GeV.
}} 
\label{fig:posA0} 
\end{figure}

Guided by the upper right panel of Fig.~\ref{fig:mh}, we now consider
$(m_{1/2}, m_0)$ planes for $A_0 > 0$. Initially, we consider in Fig.~\ref{fig:posA0}
examples with fixed $A_0 = 3$~TeV (except for the lower right panel, where $A_0 = 2$~TeV)
and $\tan \beta = 10$ (upper left panel), 40 (upper right panel), and 55 (lower panels)~\footnote{The true WMAP strips
corresponding to $\Omega_\chi h^2 = 0.112 \pm 0.012$~\cite{wmap} at the $2-\sigma$ level
are often invisibly narrow. Accordingly, in
these and most subsequent figure panels, the WMAP strips have been made more visible by colouring
regions where $0.05 < \Omega_\chi h^2 < 0.15$.}. When $\tan \beta = 10$ and 40,
we see brown shaded regions at low $m_{1/2}$ and $m_0$ where the LSP is the
lighter stop, ${\tilde t}_1$, which expand with increasing $A_0$.
There are ${\tilde t}_1 - \chi$ coannihilation strips running close to their outer
boundaries, portions of which are compatible
with $m_h \sim 119$~GeV when $\tan \beta = 10$. However, this coannihilation
region is excluded by $b \to s \gamma$ for $\tan \beta = 40$. 
In the $\tan \beta = 10$ case there is also a portion
of the ${\tilde \tau}_1 - \chi$ coannihilation strip at $m_{1/2} \sim 700$~GeV that is compatible
with $m_h \sim 119$~GeV, but no region with $m_h \sim 125$~GeV can be seen. 
On the other hand, when $\tan \beta = 40$ and $A_0 = 3$~TeV, we see that there is a portion
of the ${\tilde \tau}_1 - \chi$ coannihilation strip around $m_{1/2} \sim 800$~GeV that is compatible with $m_h = 125$~GeV.
When $\tan \beta = 55$ (lower panels of Fig.~\ref{fig:posA0}),
the ${\tilde t}_1 - \chi$ coannihilation strips disappear, and the ${\tilde \tau}_1 - \chi$
coannihilation strip morphs into the $H/A$ rapid-annihilation funnel for $m_{1/2} \sim 1500$~GeV~\footnote{Also
visible in these panels between $m_{1/2} \sim 1000$~GeV and $\sim 1500$~GeV is another WMAP-compatible
strip running roughly parallel to the ${\tilde \tau}_1 - \chi$ coannihilation strip, which is due to rapid
${\tilde \tau}_1 - {\bar{\tilde \tau}_1}$ annihilation through direct-channel $H/A$ poles.}.
In both the cases $A_0 = 3000$~GeV (lower left panel) and $A_0 = 2000$~GeV
(lower right panel), in the funnel regions
there are portions of the WMAP-compatible strips that are
compatible with $m_h = 125$~GeV, within the expected {\tt FeynHiggs}
uncertainty of $\pm 1.5$~GeV. These examples confirm that larger values of
$\tan \beta \sim 40$ or more and $A_0 > 0$ would be favoured if $m_h = 125$~GeV,
as already suggested by the upper right panel of Fig.~\ref{fig:mh} and the right panel of Fig.~\ref{fig:strips}.
Finally, we note that there is no $g_\mu - 2$-friendly region in any panel of Fig.~\ref{fig:posA0}.

\begin{figure}
\vskip 0.5in
\vspace*{-0.75in}
\begin{minipage}{8in}
\epsfig{file=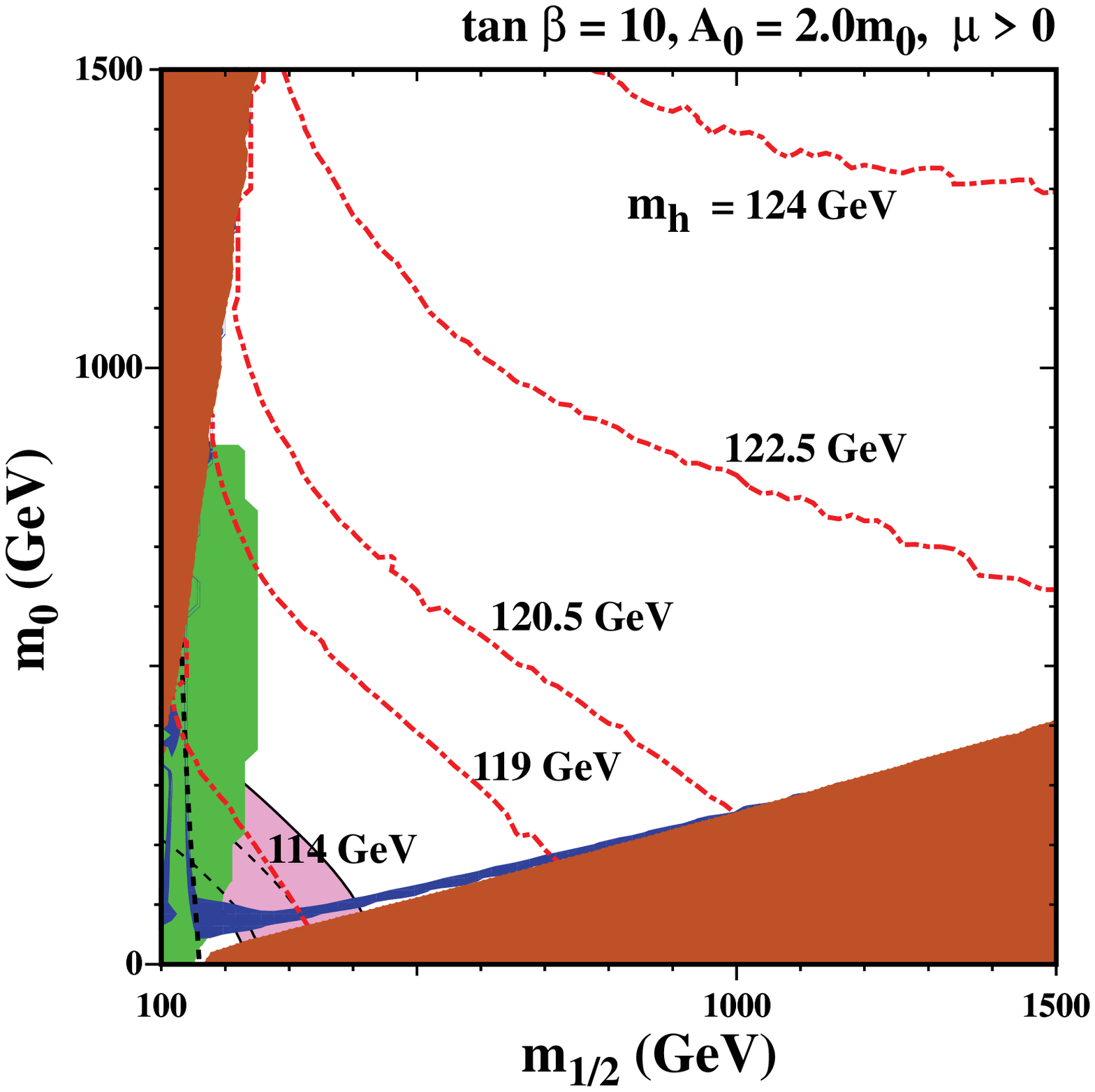,height=3.3in}
\hspace*{-0.17in}
\epsfig{file=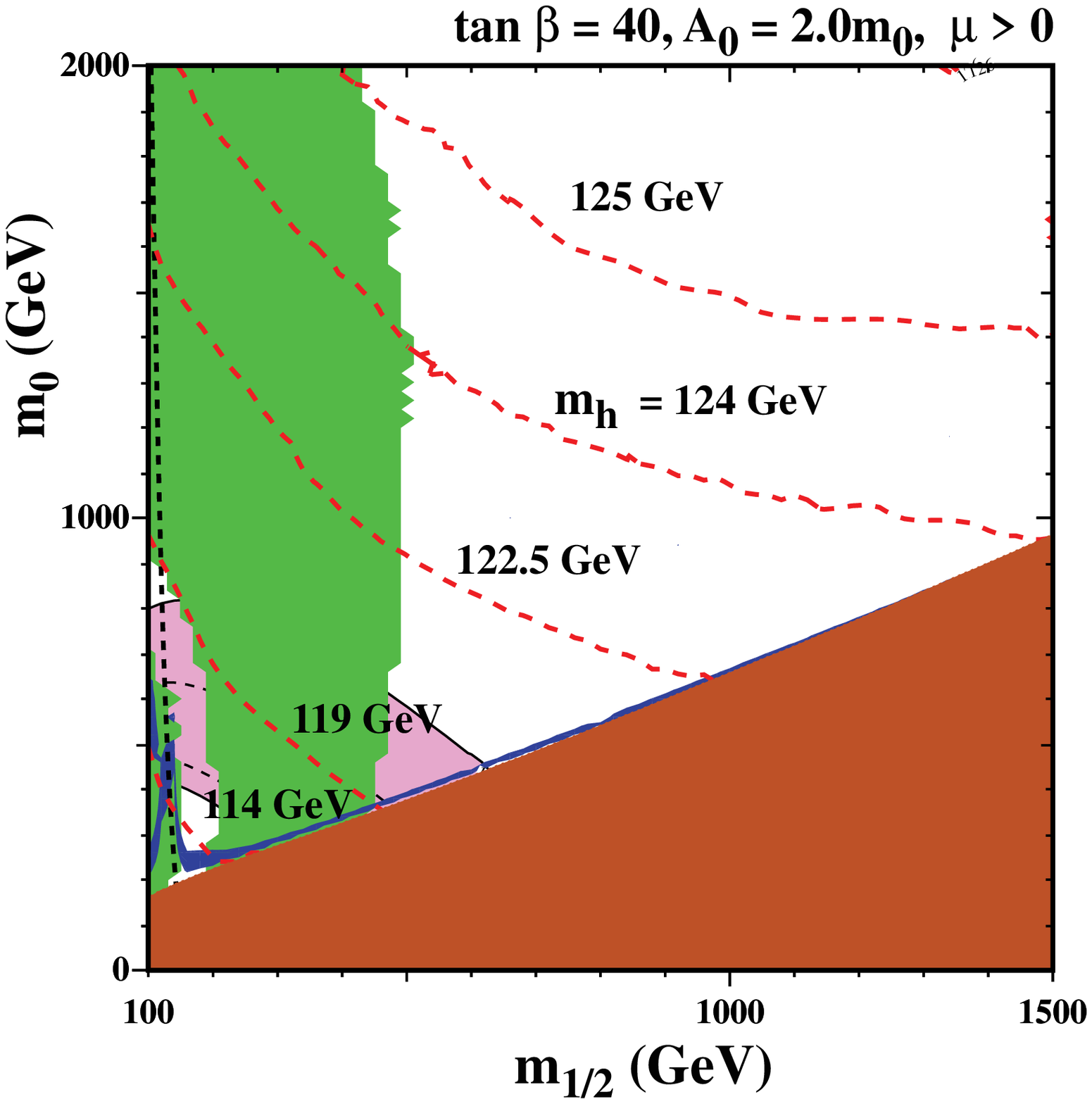,height=3.3in}
\hfill
\end{minipage}
\begin{minipage}{8in}
\epsfig{file=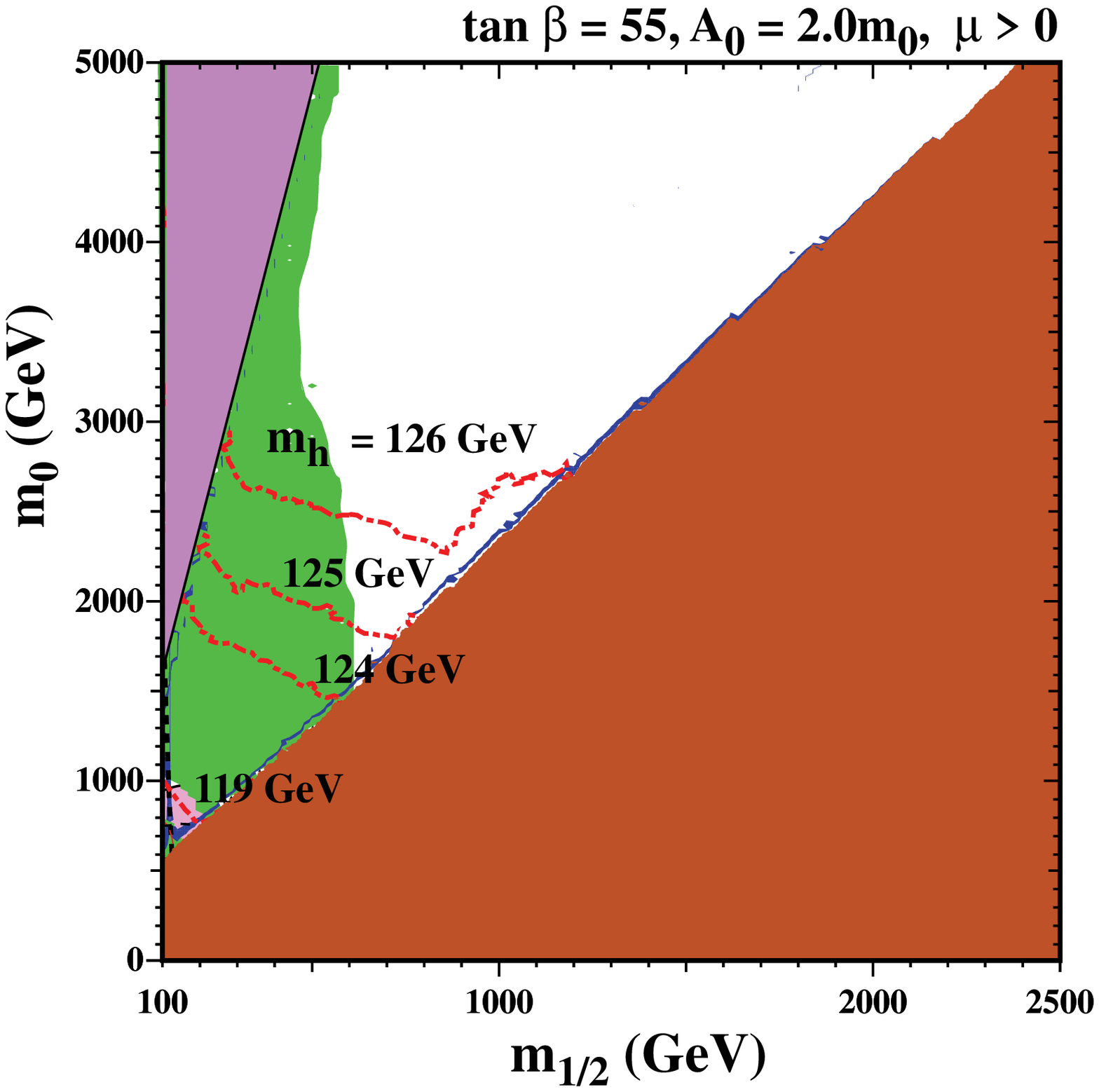,height=3.3in}
\hspace*{-0.17in}
\epsfig{file=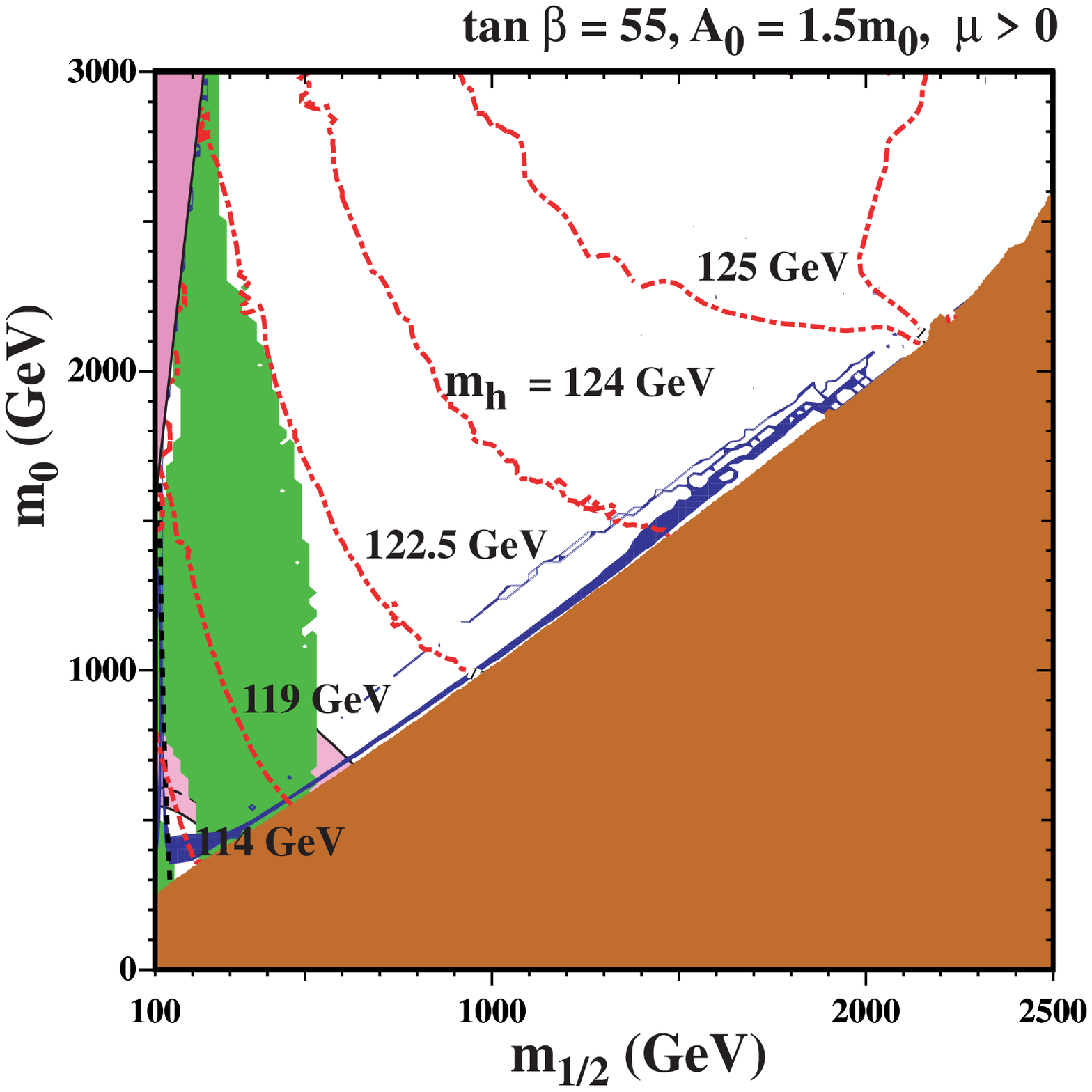,height=3.3in}
\hfill
\end{minipage}
\caption{
{\it
As Fig.~\protect\ref{fig:strips}, with $\tan \beta = 10$ (upper left panel),
 $\tan \beta = 40$ (upper right panel) and $\tan \beta = 55$ in the lower panels.
 We choose $A_0 = 2 m_0$, except in the lower right panel where $A_0 = 1.5 m_0$.
}}
\label{fig:m0A0} 
\end{figure}

Simple supergravity models of soft supersymmetry breaking suggest
a relation between $A_0$ and $m_0$ of the form $A_0 = c.m_0$ for
some constant $c \in [-3, 3]$. In this context, the fixed values of $A_0$
chosen in Fig.~\ref{fig:posA0} might appear quite extreme for small values
of $m_0$, e.g., near the junction of the ${\tilde \tau_1}$ and ${\tilde t_1}$
coannihilation strips in the upper left panel, where we find $m_h \sim 124$~GeV for $m_0 \sim 350$~GeV. 
Therefore, we display in Fig.~\ref{fig:m0A0} 
some examples of $(m_{1/2}, m_0)$ planes for fixed ratios
$A_0/m_0 = 2$ (upper panels and lower left panel) and 1.5 (lower right panel). The upper left panel
is for $\tan \beta = 10$, the upper right for $\tan \beta = 40$, and the lower panels for
$\tan \beta = 55$. As $\tb$ increases for fixed $A_0/m_0 = 2$, we see that the contours
of $m_h$ move towards smaller values of $(m_{1/2}, m_0)$, whereas the region
disallowed by $b \to s \gamma$ expands to larger $(m_{1/2}, m_0)$.
When $\tb = 10$ (upper left panel of Fig.~\ref{fig:m0A0}), we note that
there is a forbidden ${\tilde t}_1$ LSP region at small $m_{1/2}$ and large $m_0$,
which disappears for $\tb = 40$, and is replaced for $\tb = 55$
by a region where electroweak symmetry breaking is absent. The focus-point
strip adjacent to the boundary of this region is forbidden by $b \to s \gamma$
out to larger values of $m_0$ and $m_{1/2}$ than those shown.
In the lower right panel, the coannihilation strip extends into a rapid-annihilation
funnel compatible with $m_h = 125$~GeV, and we see again a rapid ${\tilde \tau}_1 -  {\bar{\tilde \tau}_1}$ annihilation
strip. 

When $\tb = 10$, there is a portion of
the ${\tilde \tau_1} - \chi$ coannihilation strip with $m_{1/2} \sim 700$~GeV that
is compatible with $m_h = 119$~GeV, and also a portion of the ${\tilde t}_1$
coannihilation strip with $m_0 \sim 1000$~GeV, but no visible region allowed by WMAP is
compatible with $m_h = 125$~GeV. 
When $\tb = 40$, the portion of the
${\tilde \tau_1} - \chi$ coannihilation strip compatible with $m_h = 119$~GeV
moves down to $m_{1/2} \sim 500$~GeV, and is one of the few cases compatible
with $g_\mu - 2$, but $m_h = 125$~GeV is still not allowed. When $\tb = 55$
and $A_0 = 2 m_0$ there is a substantial stretch of the ${\tilde \tau_1} - \chi$ coannihilation strip 
that is compatible with $m_h = 125$~GeV. When $\tb = 55$ and $A_0 = 1.5 m_0$
(lower right panel), the values of $m_h$ are generally reduced, but $m_h = 125$~GeV
is still possible in the rapid-annihilation funnel extension of the ${\tilde \tau_1} - \chi$
coannihilation strip at $m_{1/2} \sim 2000$~GeV, and $m_h = 119$~GeV is possible
for $m_{1/2} \sim 600$~GeV, in a portion of the ${\tilde \tau_1} - \chi$ coannihilation strip 
that is compatible with both $b \to s \gamma$ and $g_\mu - 2$. In this case, WMAP becomes
compatible with  $m_h = 125$~GeV along the focus point strip  when $m_0 > 5000$ GeV.

We note that when $\tb = 40$ or 55 the region forbidden by $b \to s \gamma$ is split
in two parts at smaller and larger $m_{1/2}$, separated by a strip that is allowed.
This occurs because BR($b \to s \gamma$) is too large at small $m_{1/2}$, falls through
the acceptable range as $m_{1/2}$ increases, becoming unacceptably small because of
cancellations over a range of $m_{1/2}$, before rising towards the Standard Model value
at large $m_{1/2}$. Portions of the $b \to s \gamma$-compatible band are compatible
with the cold dark matter density and/or $g_\mu - 2$, and there are also small ranges
of parameters where $m_h \sim 119$~GeV is possible. 

\begin{figure}
\vskip 0.5in
\vspace*{-0.75in}
\begin{minipage}{8in}
\epsfig{file=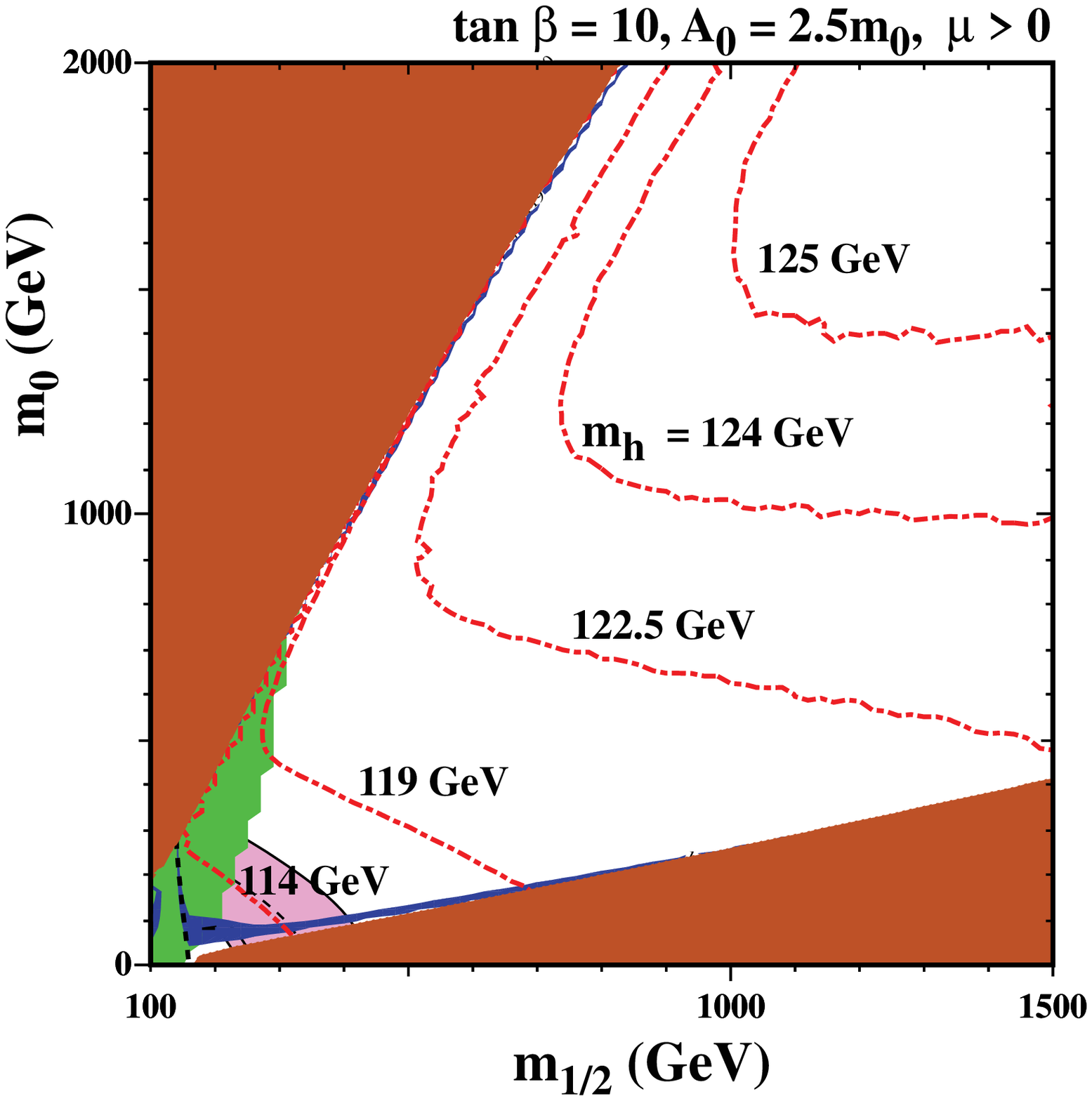,height=3.3in}
\hspace*{-0.17in}
\epsfig{file=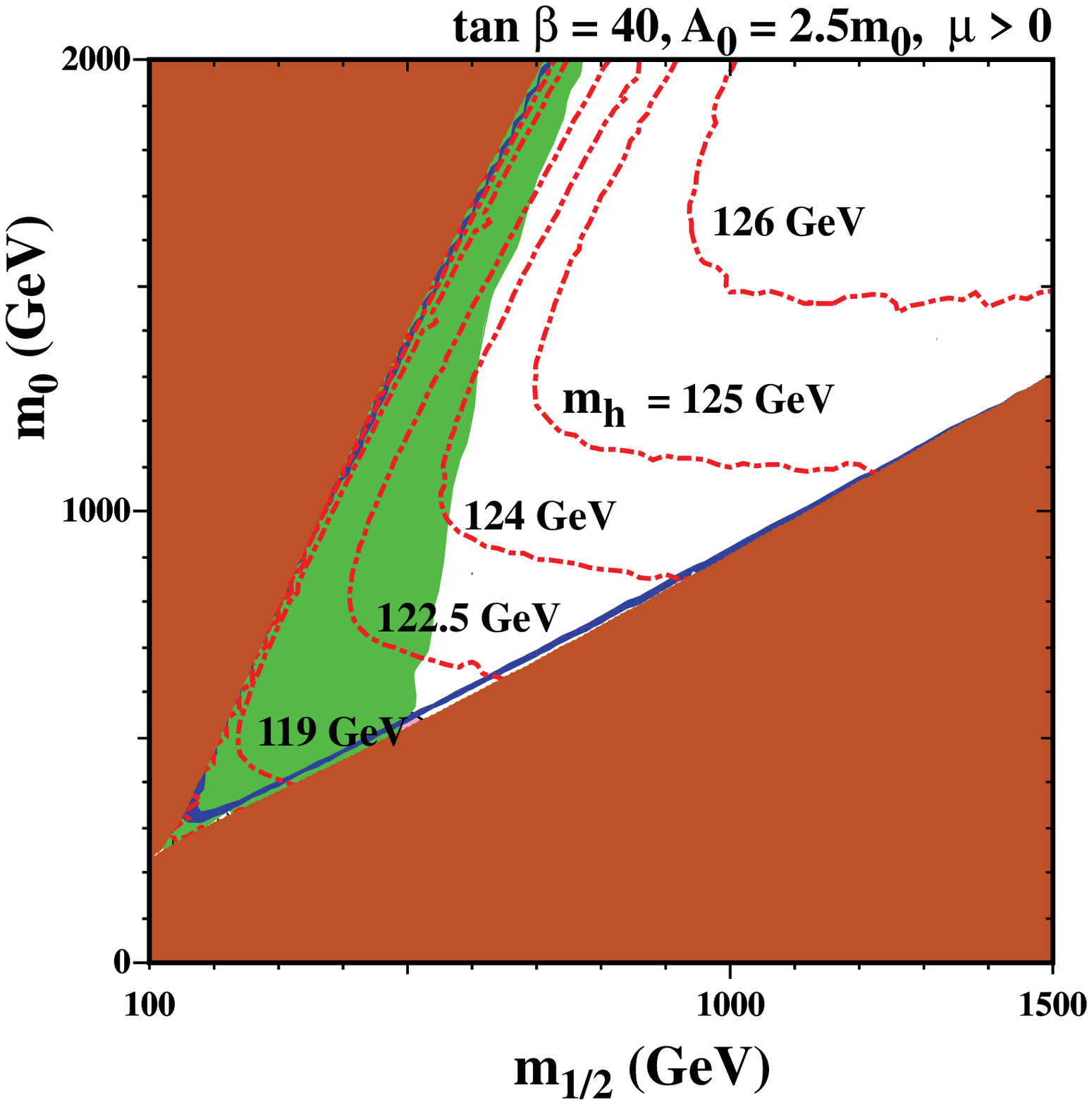,height=3.3in}
\hfill
\end{minipage}
\caption{
{\it
As Fig.~\protect\ref{fig:strips}, with $\tan \beta = 10$,  ( left panel) and
 $\tan \beta = 40$ (right panel), for $A_0 = 2.5 m_0$.}} 
\label{fig:2.5m_0} 
\end{figure}

In Fig.~\ref{fig:2.5m_0} we display similar $(m_{1/2}, m_0)$ planes for
$\tb = 10$ (left panel) and $\tb = 40$ (right panel), both with larger $A_0/m_0 = 2.5$.
In these cases, we see expanded ${\tilde t}_1$ LSP regions at large $m_0$. In the $\tb = 10$
case, $m_h = 119$~GeV is possible in portions of both the ${\tilde \tau}_1 - \chi$ and
${\tilde t}_1 - \chi$ coannihilation strips. In the $\tb = 40$ case, $m_h = 125$~GeV is possible
at large $m_{1/2}$ along the ${\tilde \tau}_1 - \chi$ coannihilation strip, but $b \to s \gamma$
forbids lower $m_h \sim 119$~GeV, and also excludes the visible part of the
${\tilde t}_1 - \chi$ coannihilation strip. For $\tb = 55$ and $A_0/m_0 = 2.5$, 
we do not find consistent solutions for generic regions of the $(m_{1/2}, m_0)$ plane.

\begin{figure}
\vskip 0.5in
\vspace*{-1.1in}
\begin{minipage}{8in}
\epsfig{file=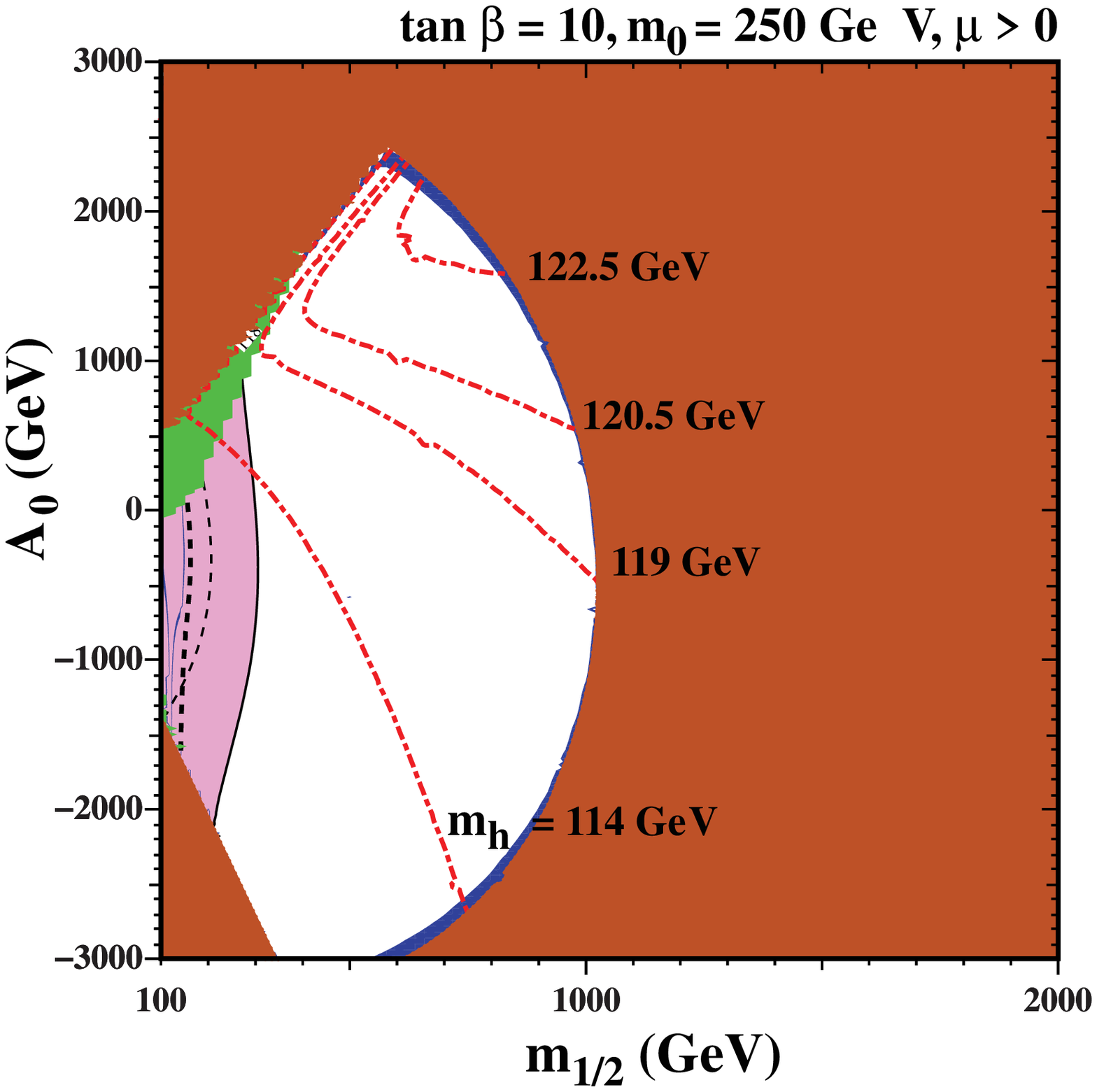,height=3.1in}
\hspace*{-0.17in}
\epsfig{file=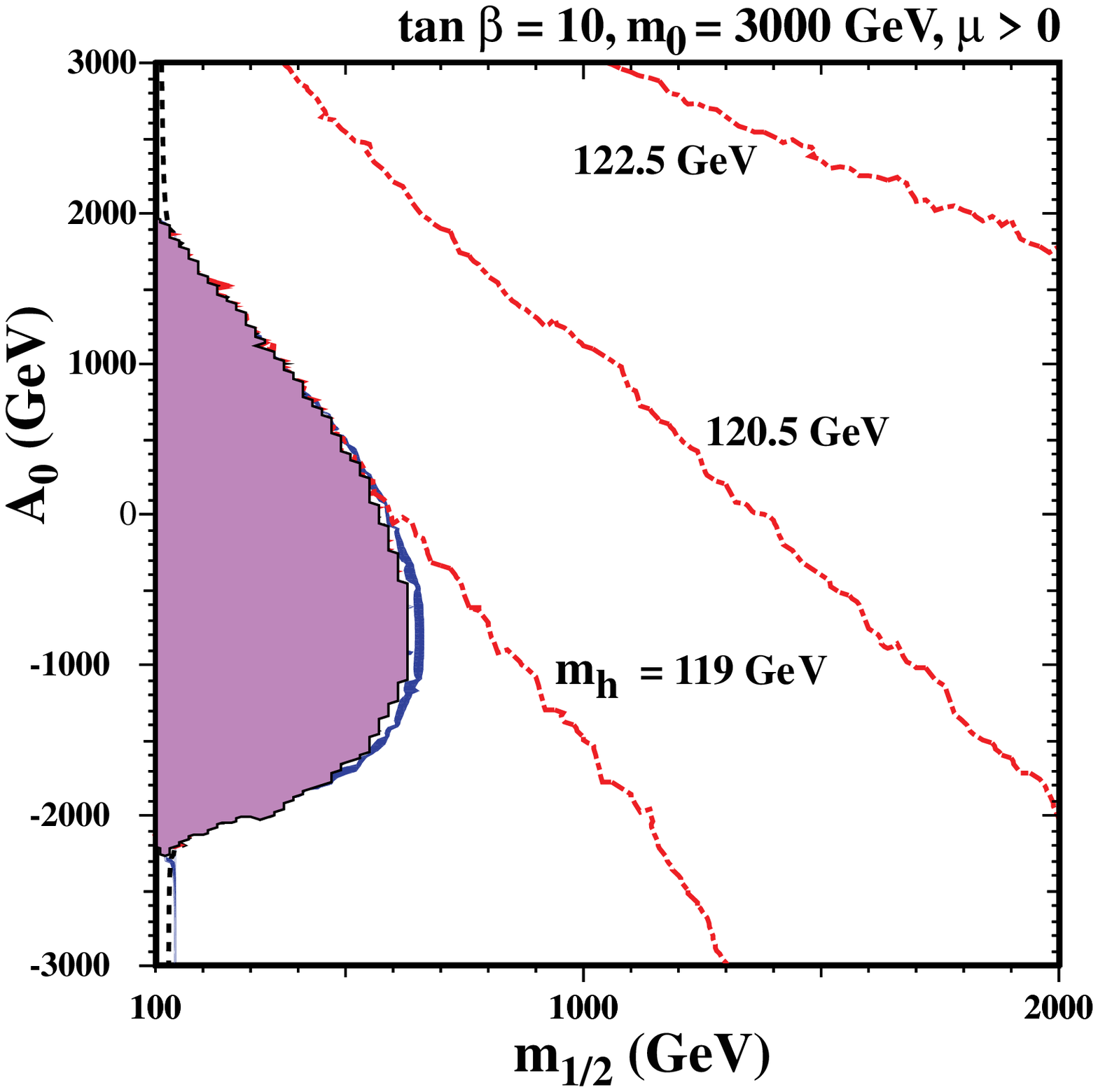,height=3.1in}
\hfill
\end{minipage}
\begin{minipage}{8in}
\epsfig{file=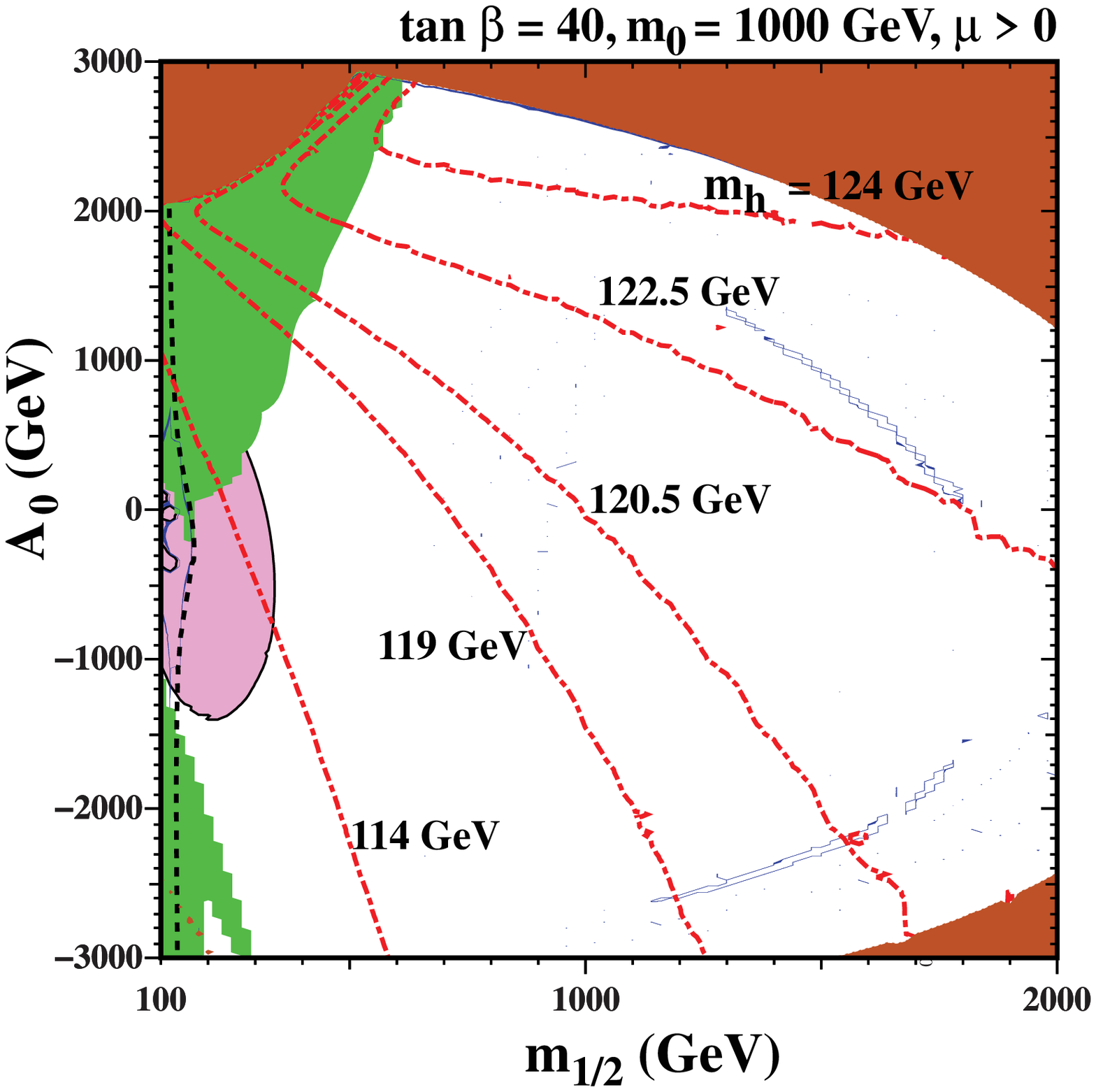,height=3.1in}
\hspace*{-0.17in}
\epsfig{file=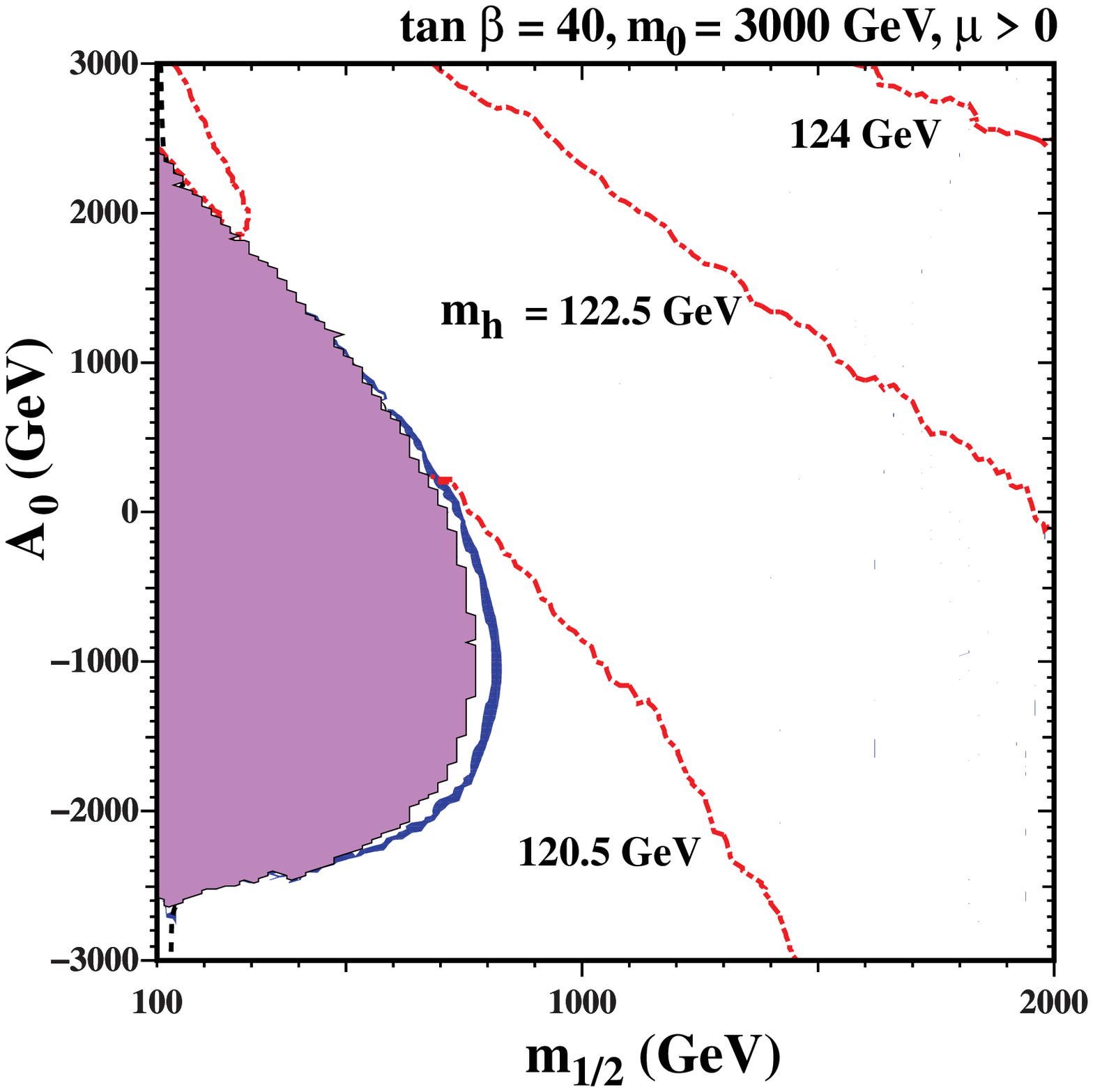,height=3.1in}
\hfill
\end{minipage}
\begin{minipage}{8in}
\epsfig{file=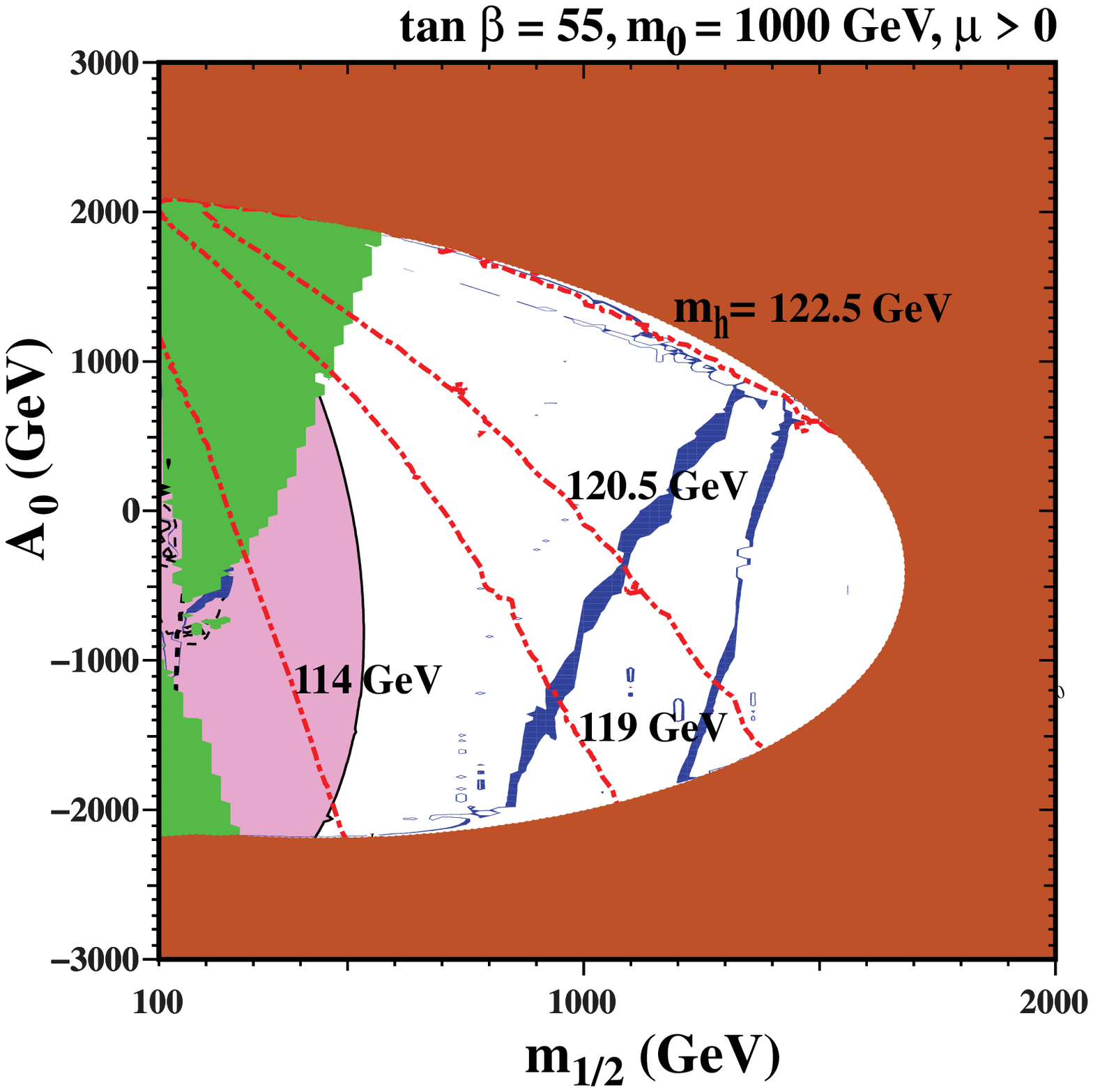,height=3.1in}
\hspace*{-0.17in}
\epsfig{file=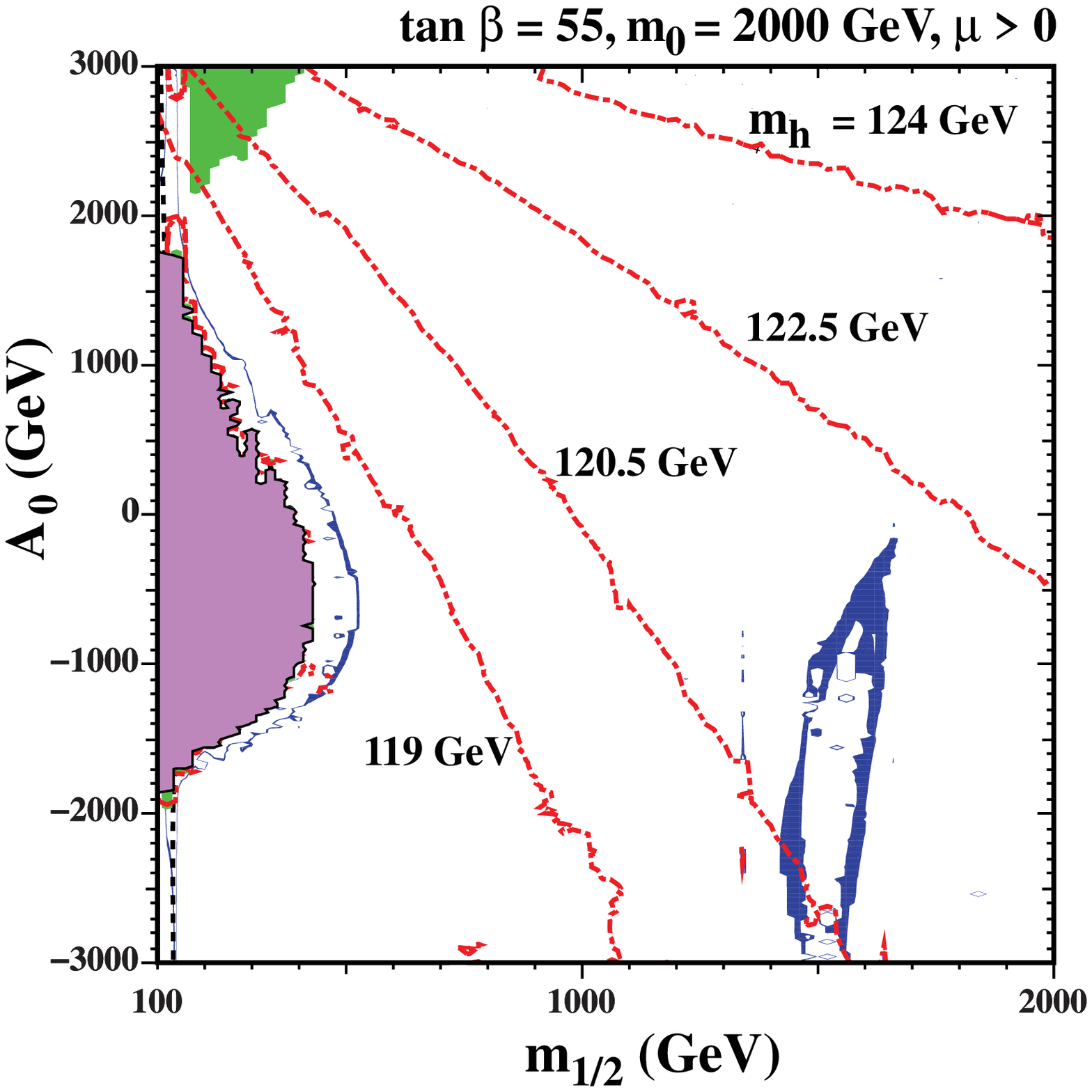,height=3.1in}
\hfill
\end{minipage}
\caption{
{\it
Representative $(m_{1/2}, A_0)$ planes for $\mu > 0$, $\tan \beta = 10$
with $m_0 = 250$~GeV (top left panel) and $3000$~GeV (top right panel),
or $\tan \beta = 40$
with $m_0 = 1000$~GeV (middle left panel) and $3000$~GeV (middle right panel),
and for $\tan \beta = 55$
with $m_0 = 1000$~GeV (bottom left panel) and $2000$~GeV (bottom right panel),
with the same shadings and contours as previously.}} 
\label{fig:m12A0} 
\end{figure}

In order to see from a different perspective the influence of the choice of $A_0$,
in Fig.~\ref{fig:m12A0} we display $(m_{1/2}, A_0)$ planes for
$\tb = 10$ (top panels), $\tb = 40$ (middle panels), $\tb = 55$ (bottom panels), and
low values of $m_0 = 250, 1000, 1000$~GeV (left panels) compared with 
the large values $m_0 = 3000, 3000, 2000$~GeV (right panels). In the top left
panel for $\tb = 10$ we see a large ${\tilde \tau}_1$ LSP region at large $m_{1/2}$
and two ${\tilde t}_1$ LSP regions at small $m_{1/2}$ and large $|A_0|$.
Adjacent to the boundaries of these regions there are coannihilation strips,
and inside the $\chi$ LSP region there is a region in conflict with
$b \to s \gamma$ and a region favoured by $g_\mu - 2$. We see clearly
that $m_h$ increases with increasing $A_0$. There is no
portion of the allowed region that is compatible with $m_h = 125$~GeV,
but there are two WMAP-compatible regions with $m_h \sim 119$~GeV:
along the ${\tilde t}_1 - \chi$ coannihilation strip for $A_0 > 0$ and on the
${\tilde \tau}_1 - \chi$ coannihilation strip where $m_{1/2} \sim 1000$~GeV
and $A_0 \sim - 500$~GeV.
As $m_0$ is increased, the ${\tilde \tau}_1$ LSP region recedes to large $m_{1/2}$
and the ${\\\tilde t}_1$ LSP regions recede to larger $|A_0|$. At the same time, the
regions excluded by $b \to s \gamma$ and the $g_\mu - 2$-compatible region
disappear from the visible area of the $(m_{1/2}, A_0)$ plane. Instead, in the
top right panel of Fig.~\ref{fig:m12A0} for $\tb = 10$ and $m_0 = 3000$~GeV, we
see the appearance of a protuberance where there is no consistent electroweak
symmetry breaking, which is surrounded by a focus-point strip. Taking into account
the theoretical uncertainties, all the displayed portion of this strip is compatible with 
$m_h \sim 119$~GeV, with $A_0 > 0$ preferred.

Examples with $\tb = 40$ are shown in the middle panels of Fig.~\ref{fig:m12A0}.
For $m_0 = 1000$~GeV (left panel), we again see ${\tilde \tau}_1$ and ${\tilde t}_1$
regions flanked by coannihilation strips, which have receded from their locations in the top left panel for
$\tb = 10$ and $m_0 = 250$~GeV. We also see a $g_\mu - 2$-compatible region at
small $m_{1/2}$, and note that the region forbidden by $b \to s \gamma$ has expanded
as compared with the top left panel. The only WMAP compatible region where $m_h \sim 125$~GeV
is at $m_{1/2} \sim 500$ to 1500~GeV with $A_0 > 2000$~GeV, whereas $m_h \sim 119$~GeV
is possible only for $A_0 < -3000$~GeV. Turning to the case $\tb = 40$, $m_0 = 3000$~GeV
(middle right panel of Fig.~\ref{fig:m12A0}), we see that the protuberance without electroweak
symmetry breaking has expanded. In this case all the surrounding focus-point strip is
compatible with $m_h \sim 119$~GeV, but $m_h = 125$~GeV is out of reach.

Turning finally to the bottom panels of Fig.~\ref{fig:m12A0} for $\tb = 55$, we
see new features, namely rapid-annihilation funnels centred around $m_{1/2} \sim 1000$~GeV
if $m_0 = 1000$~GeV and around $m_{1/2} \sim 1500$~GeV if $m_0 = 2000$~GeV~\footnote{These
funnels have been coloured only in the range $\Omega_\chi h^2 = 0.112 \pm 0.012$ allowed
by WMAP at the $2-\sigma$ level.}.
In both cases there are portions of the funnel regions with $A_0 < 0$ that
are also compatible with $m_h \sim 119$~GeV, as well as portions of the ${\tilde \tau}_1$
coannihilation strip with $A_0 < 0$ for $m_0 = 1000$~GeV and the $A> 0$ portion of the
focus-point strip for $m_0 = 2000$~GeV, but $m_h = 125$~GeV is again nowhere to be seen.

\section{WMAP Strips for $\tb = 10, 40$ and $55$}

In light of the above illustrations, we will consider the detectability of neutralino dark
matter along some characteristic WMAP strips in the CMSSM parameter space,
paying particular attention to examples where $m_h \sim 119$ or 125~GeV.

We first consider examples with $\tb = 10$. Comparing the left panel of Fig.~\ref{fig:strips}
with the upper left panels of Figs.~3 and 4, and the left panel of Fig. 5, we see that the
values of $m_h$ along the ${\tilde \tau}_1 - \chi$ coannihilation strip increase only slowly
with $A_0/m_0$. Accordingly, in the following we consider this strip in the cases
$A_0 = 0$ and $A_0 = 2.5 m_0$. These are parametrized approximately by
(here and in the following equations, dimensionful parameters are expressed in GeV units):
\begin{eqnarray}
m_0 & = & 0.24 \  m_{1/2} - 0.49 , \qquad A_0 = 0 ,  \label{10stau0}\\
m_0 & = & 0.25 \ m_{1/2} + 3.50  , \qquad A_0 = 2.5 m_0 .
\label{10stau2.5}
\end{eqnarray}
Values of $m_h \sim 119$~GeV are attained for relatively large values of $m_{1/2}$
along these strips. For comparison, we also discuss the focus-point strip for $\tb = 10$
and $A_0 = 0$, which may be parametrized (a linear fit is inadequate in this case) by
\begin{eqnarray}
m_0 & = & - 0.0011 \  m_{1/2}^2 + 4.50 \  m_{1/2} + 750 ,
\label{10fp}
\end{eqnarray}
much of which is compatible with $m_h = 119$~GeV.

Several figures exhibit ${\tilde t}_1 - \chi$ coannihilation strips, which
have not been much discussed in the dark matter detection literature, so we choose one
example of this possibility. Specifically, the ${\tilde t}_1 - \chi$ strip in the left panel of 
Fig.~5 for $\tb = 10$ and $A_0 = 2.5 m_0$ is parametrized approximately by
\begin{eqnarray}
m_0 & = & 2.40 \ m_{1/2} - 14 .
\label{10stop}
\end{eqnarray}
Points all along this strip give values of $m_h$ consistent with 119~GeV, taking into consideration
the theoretical uncertainties in the calculation of $m_h$.

We also consider the corresponding ${\tilde \tau}_1 - \chi$ coannihilation strips for $\tb = 40$,
for $A_0 = 0$ and $2.5m_0$:
\begin{eqnarray}
m_0 & = & 0.34 \ m_{1/2} + 82 , \qquad A_0 = 0 ,\label{40stau0} \\
m_0 & = & 0.75 \ m_{1/2} + 160 , \qquad A_0 = 2.5 m_0 .
\label{40stau2.5}
\end{eqnarray}
We also consider some examples of strips with $\tb = 55$.
In the case $A_0 = 0$, shown in the right panel of Fig.~\ref{fig:strips}, the
${\tilde \tau}_1 - \chi$ coannihilation strip morphs into the rapid $H/A$ annihilation
funnel when $m_{1/2} \sim 1000$~GeV, so we give parametrizations of both sides
of the funnel:
\begin{eqnarray}
m_0 & = & 0.0010 \  m_{1/2}^2 - 0.49 \ m_{1/2} + 390 , \nonumber \\
m_0 & = & 3.6 \ m_{1/2} -3700 .
\label{55stau0}
\end{eqnarray}
In this case, $m_h \sim 119$~GeV at relatively low values of $m_{1/2}$ below the
funnel bifurcation, but values compatible with $m_h = 125$~GeV are not reached
even at the tip of the funnel.
As already commented, when $\tb = 55$ we do not find generic solutions for
$A_0 = 2.5 m_0$, so we consider $A_0 = 2 m_0$, as shown in the lower left
panel of Fig.~\ref{fig:m0A0}. This strip does not bifurcate into a funnel, and is parametrized by
\begin{eqnarray}
m_0 & = & 1.9 \ m_{1/2} + 430 .
\label{55stau2.0}
\end{eqnarray}
In this case, values of $m_h \sim 125$~GeV and even larger are quite possible.
Finally, we consider the focus-point strip for $\tb = 55$ and $A_0 = 0$:
\begin{eqnarray}
m_0 & = & - 0.0011 \ m_{1/2}^2 + 3.8 \ m_{1/2} + 290 ,
\label{55fp}
\end{eqnarray}
where (we recall) all dimensionful parameters in the above equations are expressed in GeV units.

\section{The Higgs Mass and Dark Matter Scattering along WMAP Strips}

We now discuss the the spin-independent
dark matter scattering cross section along the WMAP strips introduced in the
previous Section, and correlate it with the predicted mass of the Higgs boson. 

\begin{figure}[htb!]
\vskip 0.5in
\vspace*{-0.75in}
\begin{minipage}{8in}
\epsfig{file=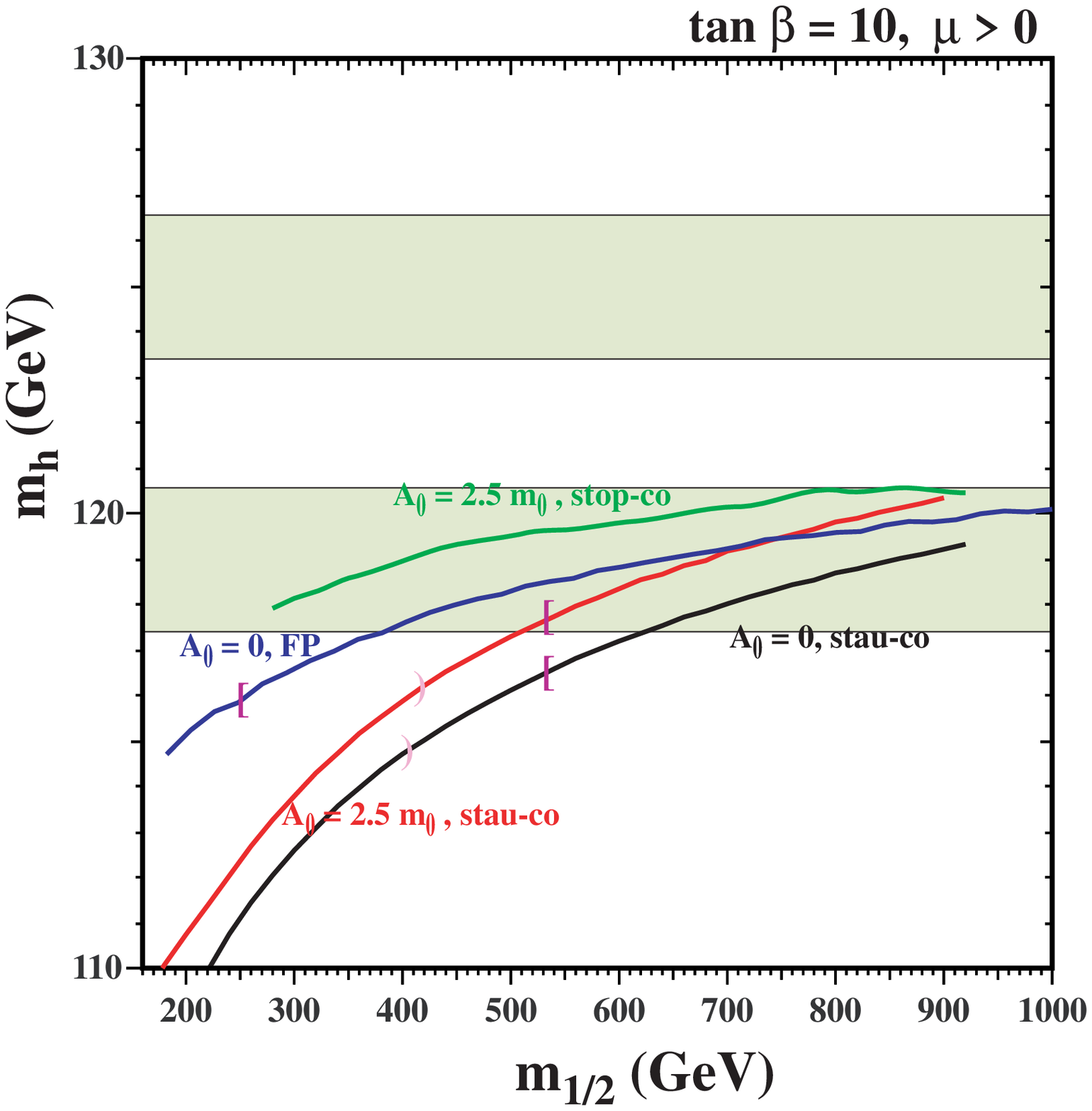,height=3in}
\epsfig{file=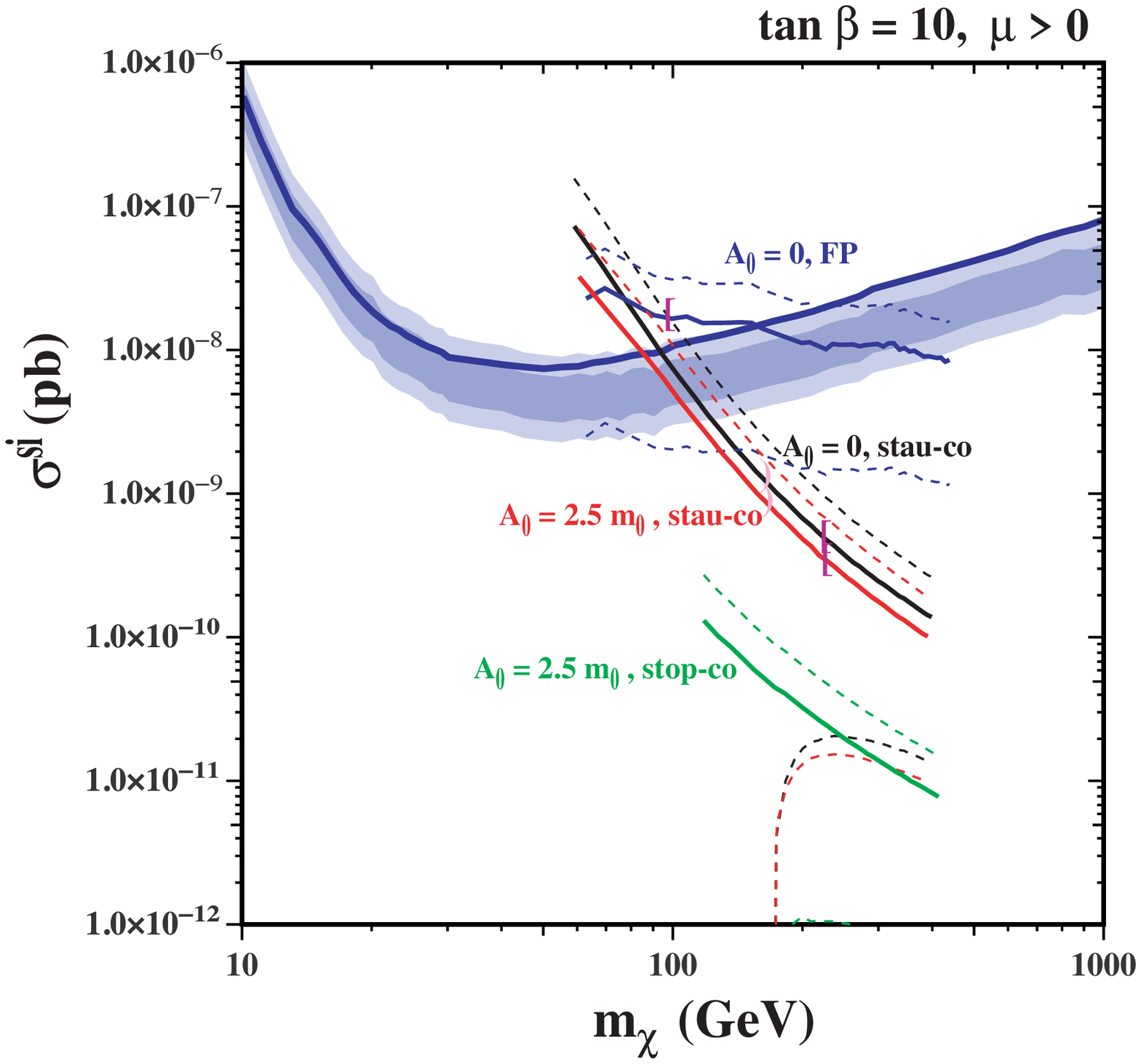,height=3in}
\hfill
\end{minipage}
\vspace{0.1in}
\caption{
{\it
Left panel: $m_h$ as calculated using {\tt FeynHiggs} (showing the band $m_h = 119 \pm 1.5$~GeV)
and right panel: spin-independent elastic $\chi-p$ scattering cross section
(showing the XENON100 upper limit~\protect\cite{XENON100} 
as a solid line accompanied by the shaded band described in the text), 
along the WMAP strips for
$\tb = 10$ - the ${\tilde \tau}_1 - \chi$ coannihilation strips for $A_0 = 0$ (\protect\ref{10stau0}) 
(black) and $A_0/m_0 = 2.5$ (\protect\ref{10stau2.5}) (red),
the focus-point strip for $A_0 = 0$ (\protect\ref{10fp}) (blue),
and the ${\tilde t}_1 - \chi$ coannihilation strip for $A_0/m_0 = 2.5$ (\protect\ref{10stop}) (green).
In the left panels of this and subsequent figures, the ranges $m_h = 119 \pm 1.5$~GeV and $125 \pm 1.5$~GeV
are green shaded horizontal bands. Here and subsequently, the portions of the WMAP strips allowed by the 
LHC missing-energy searches~\protect\cite{LHCMET}
are indicated by (purple) square brackets: {\bf \large{[}} and the portions favoured by $g_\mu - 2$ are indicated by
(pink) parentheses: {\bf \large{)}}.
}}
\label{fig:10strips} 
\end{figure}

The left panel of Fig.~\ref{fig:10strips} shows $m_h$
(as calculated using {\tt FeynHiggs}) along the WMAP ${\tilde \tau}_1 - \chi$ 
coannihilation strips parametrized by (\ref{10stau0}, \ref{10stau2.5}) for $A_0 = 0$
(black line) and $A_0/m_0 = 2.5$ (red line), respectively. The upper ends of the
stau coannihilation strips at $m_{1/2} \sim 900$~GeV are where $m_\chi = m_{\tilde \tau_1}$,
and they are truncated by the LHC searches for missing-energy events at $m_{1/2} \sim 530$~GeV~\cite{LHCMET}. 
The portions of these and other strips allowed by the LHC missing-energy searches~\cite{LHCMET}
are indicated by (purple) square brackets: {\bf \large{[}} and the portions favoured by $g_\mu - 2$ are indicated by
(pink) parentheses: {\bf \large{)}}. The absence of a  {\bf \large{)}} along a line indicates that no
portion of the line is compatible with $g_\mu -2$. 
We see that these constraints are incompatible for the $\tb = 10$, $A_0 = 0$ strips shown.

As expected, increasing $A_0$
gives larger values of $m_h$, in this case by $\sim 1$~GeV, almost
independently of $m_{1/2}$. We see that $m_h$ is compatible with 119~GeV (within the
estimated {\tt FeynHiggs} error of $\pm 1.5$~GeV, indicated by the lower green shaded horizontal band) 
for $m_{1/2} > 630$~GeV for $A_0 = 0$
and $m_{1/2} = 520$~GeV for $A_0/m_0 = 2.5$. 
Also shown in this panel as a blue line is
$m_h$ along the focus-point strip for $\tb = 10$ and $A_0 = 0$ (\ref{10fp}), which is cut off below
$m_{1/2} \sim 300$~GeV by the LHC missing-energy searches~\cite{LHCMET},
and is also compatible 
within errors with 119~GeV for $m_{1/2} > 370$~GeV. Finally, the green line
shows $m_h$ along the ${\tilde t}_1 - \chi$ coannihilation strip for $\tan \beta = 10$ and $A_0/m_0 = 2.5$,
parametrized by (\ref{10stop})~\footnote{We do not indicate LHC bounds here and in other cases in which the 
available results from LHC missing-energy
searches~\cite{LHCMET} are insufficient to indicate which portions of this line might be excluded.}. We see that
$m_h$ is somewhat higher along this line, and compatible within errors with $m_h = 119$~GeV
for all the allowed range of $m_{1/2}$. On the other hand, none of these strips is compatible
with $m_h = 125$~GeV (the upper green shaded horizontal band).

The right panel of Fig.~\ref{fig:10strips} displays the spin-independent $\chi-p$
scattering cross section calculated along the same strips for $\tb = 10$, displayed as
functions of $m_\chi \sim 0.42 m_{1/2}$. The central
values (shown as solid lines) are for $\Sigma_{\pi N} = 50$~MeV,
and the dashed lines are for 64 and 36~MeV, respectively~\footnote{See~\cite{EOS} for
a discussion of the uncertainty in this parameter.}. These predictions are
compared with the upper limit from the XENON100 experiment (solid dark blue line, the
shaded bands are the ranges of the exclusion expected at the $\pm 1,2 \sigma$ levels)~\cite{XENON100}. We see that
the cross section along the  ${\tilde \tau}_1 - \chi$ coannihilation strip
for $A_0/m_0 = 2.5$ (shown in red) is somewhat lower than for 
$A_0 = 0$ (shown in black), though the
difference is much less than the hadronic uncertainty in the cross
section. If $\Sigma_{\pi N} = 50$~MeV, the portions $m_\chi < 80, 90$~GeV would be excluded by
the XENON100 experiment~\cite{XENON100}, and the LHC missing-energy searches~\cite{LHCMET} and
the hypothetical $m_h = 119$~GeV measurement would
suggest a cross section $< 10^{-9}$~pb. 
The cross section along the focus-point
strip (shown in blue) is significantly higher, particularly for large $m_\chi$.
This reflects the fact that along this strip the relic density is brought into the
WMAP range by $\chi-\chi$ annihilations alone, whereas coannihilation
processes are important along the other strips. Thus, the $\chi-\chi$
annihilation cross section is higher along this strip, and the correspondingly also
the elastic scattering cross section~\footnote{This connection is only qualitative, since the 
processes dominating $t$-channel exchange are not identical with the processes
dominating $s$-channel annihilation, and the cosmological annihilations involve a mixture
of $P-$ and $S-$wave annihilations.}. The XENON100 
experiment~\cite{XENON100} imposes $m_\chi > 150$~GeV
along this line, if $\Sigma_{\pi N} = 50$~MeV. The elastic scattering
cross section is lowest of all along the ${\tilde t}_1 - \chi$ coannihilation strip (shown in green).
This is because in this case the coannihilating partner
particle is strongly-interacting, and the weights of ${\tilde t}_1 - \chi$ coannihilations and
${\tilde t}_1 - \bar{\tilde t}_1$ and ${\tilde t}_1 - {\tilde t}_1$ annihilations are enhanced 
relative to the corresponding processes
along the ${\tilde \tau}_1 - \chi$ coannihilation strip, so the role of $\chi - \chi$ annihilation
is reduced, and similarly the elastic scattering cross section, which is far below the XENON100 upper limit.

\begin{figure}
\vskip 0.5in
\vspace*{-0.75in}
\begin{minipage}{8in}
\epsfig{file=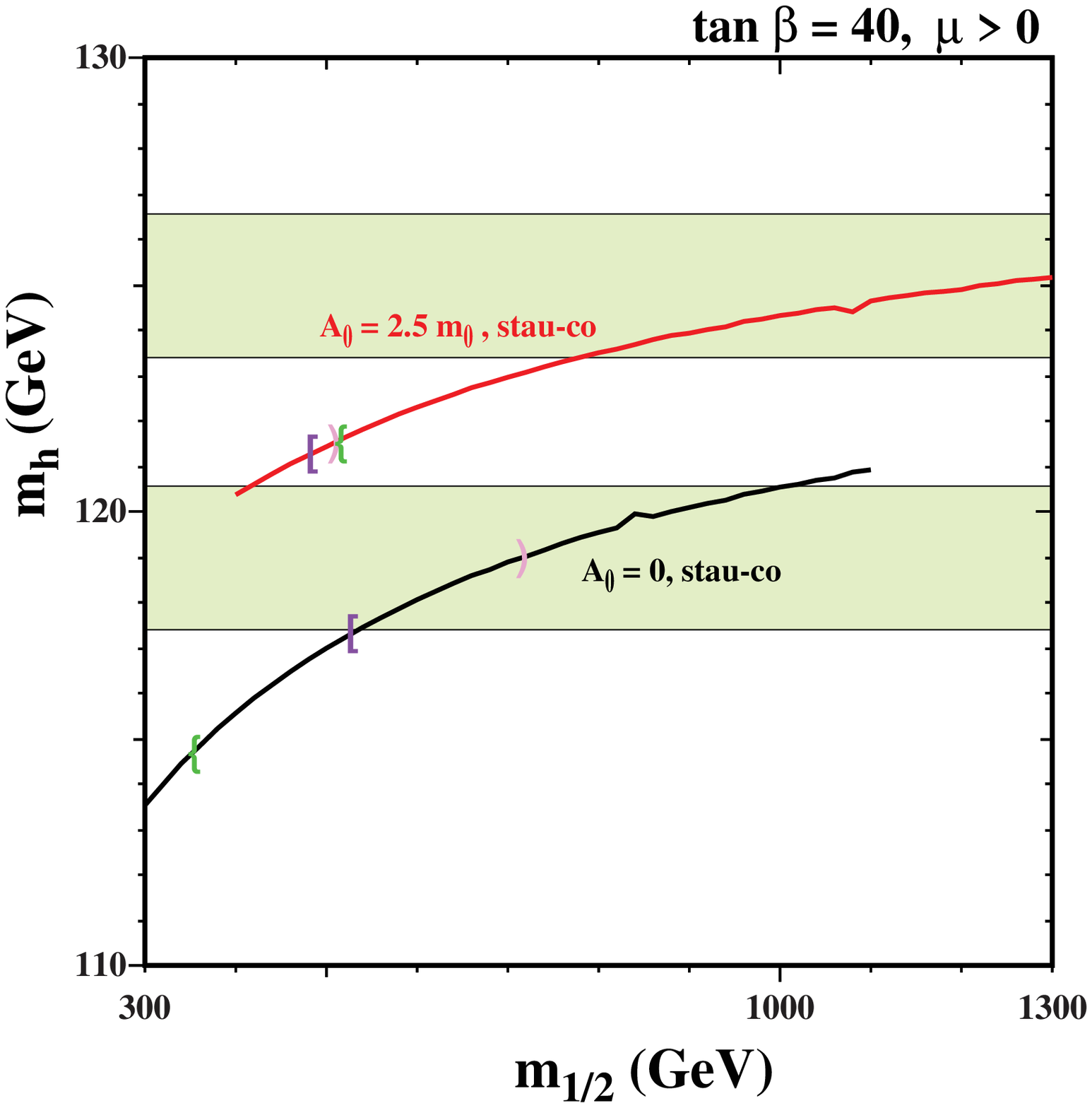,height=3in}
\epsfig{file=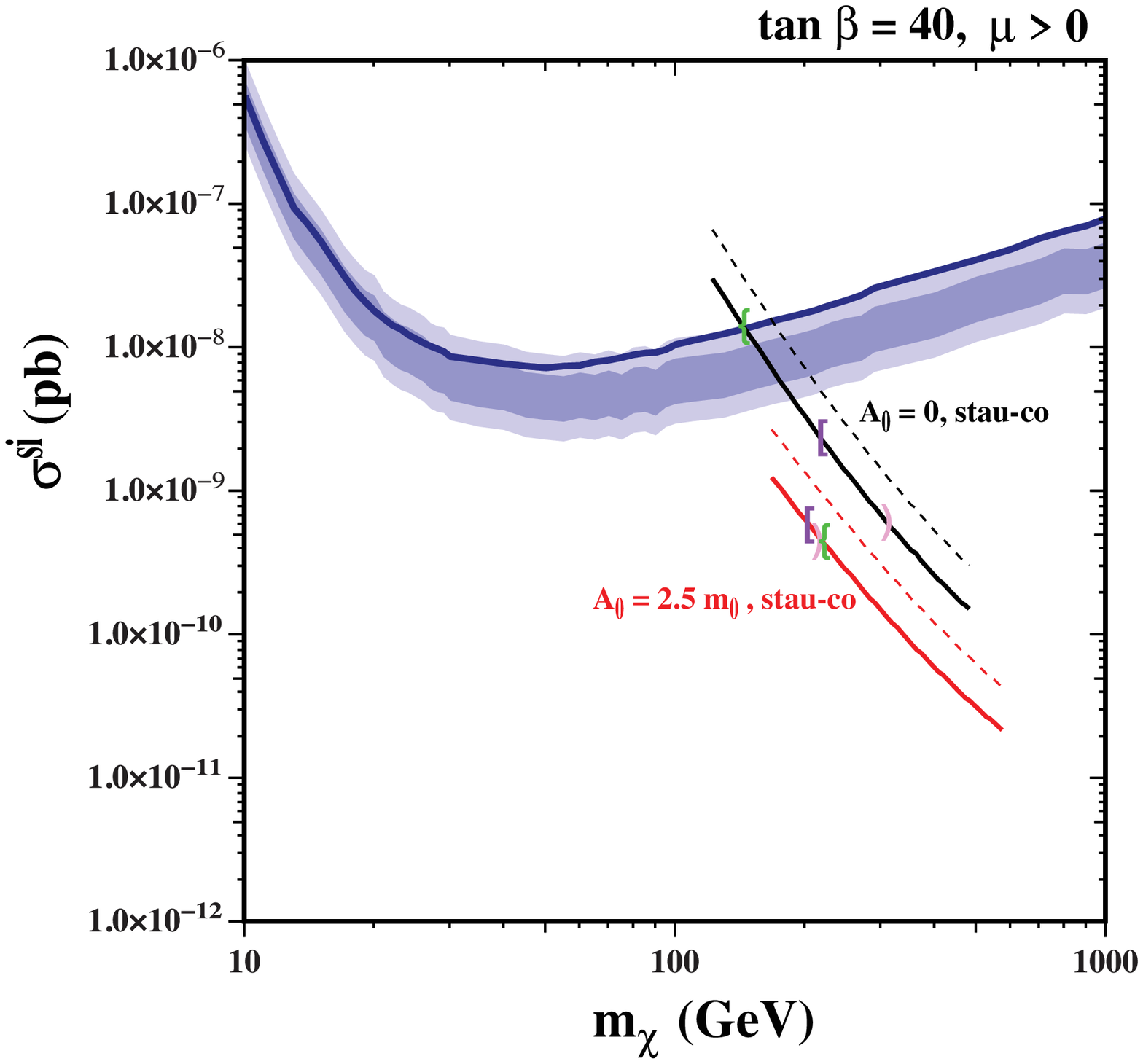,height=3in}
\hfill
\end{minipage}
\vspace{0.1in}
\caption{
{\it
Left panel: $m_h$ as calculated using {\tt FeynHiggs} (showing the bands $m_h = 119 \pm 1.5$~GeV and $125 \pm 1.5$~GeV)
and right panel: spin-independent elastic $\chi-p$ scattering cross section (showing the XENON100 exclusion~\protect\cite{XENON100}
as in Fig.~\protect\ref{fig:10strips}), along WMAP strips for
$\tb = 40$ - the ${\tilde \tau}_1 - \chi$ coannihilation strips for $A_0 = 0$ (\protect\ref{40stau0}) 
(black) and $A_0/m_0 = 2.5$ (red) (\protect\ref{40stau2.5}).
The lower bounds on $m_{1/2}$ along these strips due to $b \to s \gamma$ are indicated by green brackets  {\bf \large{\{}} .
}}
\label{fig:40strips} 
\end{figure}

The left panel of Fig.~\ref{fig:40strips} displays
the values of $m_h$ along various WMAP strips for $\tb = 40$.
As before, the black line is for the ${\tilde \tau}_1 - \chi$ coannihilation strip
with $A_0 = 0$ (\ref{40stau0}), and the (substantially higher) red line is for
the corresponding strip with $A_0/m_0 = 2.5$ (\ref{40stau2.5}). In this case, the coannihilation strip
for $A_0 = 0$ extends to $m_{1/2} \sim 1100$~GeV before terminating where $m_\chi = m_{\tilde \tau_1}$,
whereas the strip for $A_0 = 2.5$ extends to $m_{1/2} \sim 1300$~GeV.
The lower bounds on $m_{1/2}$ along these strips due to $b \to s \gamma$ are indicated by green brackets  {\bf \large{\{}} .
The $A_0 = 0$ case is compatible with $m_h = 119$~GeV
for $m_{1/2} > 550$~GeV (almost corresponding to the LHC missing-energy constraint~\cite{LHCMET}), 
and the $A_0/m_0 = 2.5$ case is compatible with $m_h = 125$~GeV
for $m_{1/2} > 700$~GeV, within the {\tt FeynHiggs} uncertainty of $\pm 1.5$~GeV, as indicated by the horizontal bands.
We see in the right panel of Fig.~\ref{fig:40strips} that the elastic scattering cross
section for $A_0/m_0 = 2.5$ (red) is smaller by almost an order of magnitude than that for
$A_0 = 0$ (black), with the (optimistic) red dashed line for $A_0/m_0 = 2.5$
with $\Sigma_{\pi N} = 64$~MeV lying below the $\Sigma_{\pi N} = 50$~MeV 
value for $A_0 = 0$. Coupled with the lower limit $m_{1/2} > 700$~MeV required
to be compatible with $m_h = 125$~GeV, we see that confirmation of this Higgs
mass would suggest a cross section below $10^{-9}$~pb in this model. On the other
hand, the XENON100 experiment~\cite{XENON100} already requires $m_\chi > 150$~GeV if $A_0 = 0$
and $\Sigma_{\pi N} = 50$~MeV.

\begin{figure}
\vskip 0.5in
\vspace*{-0.75in}
\begin{minipage}{8in}
\epsfig{file=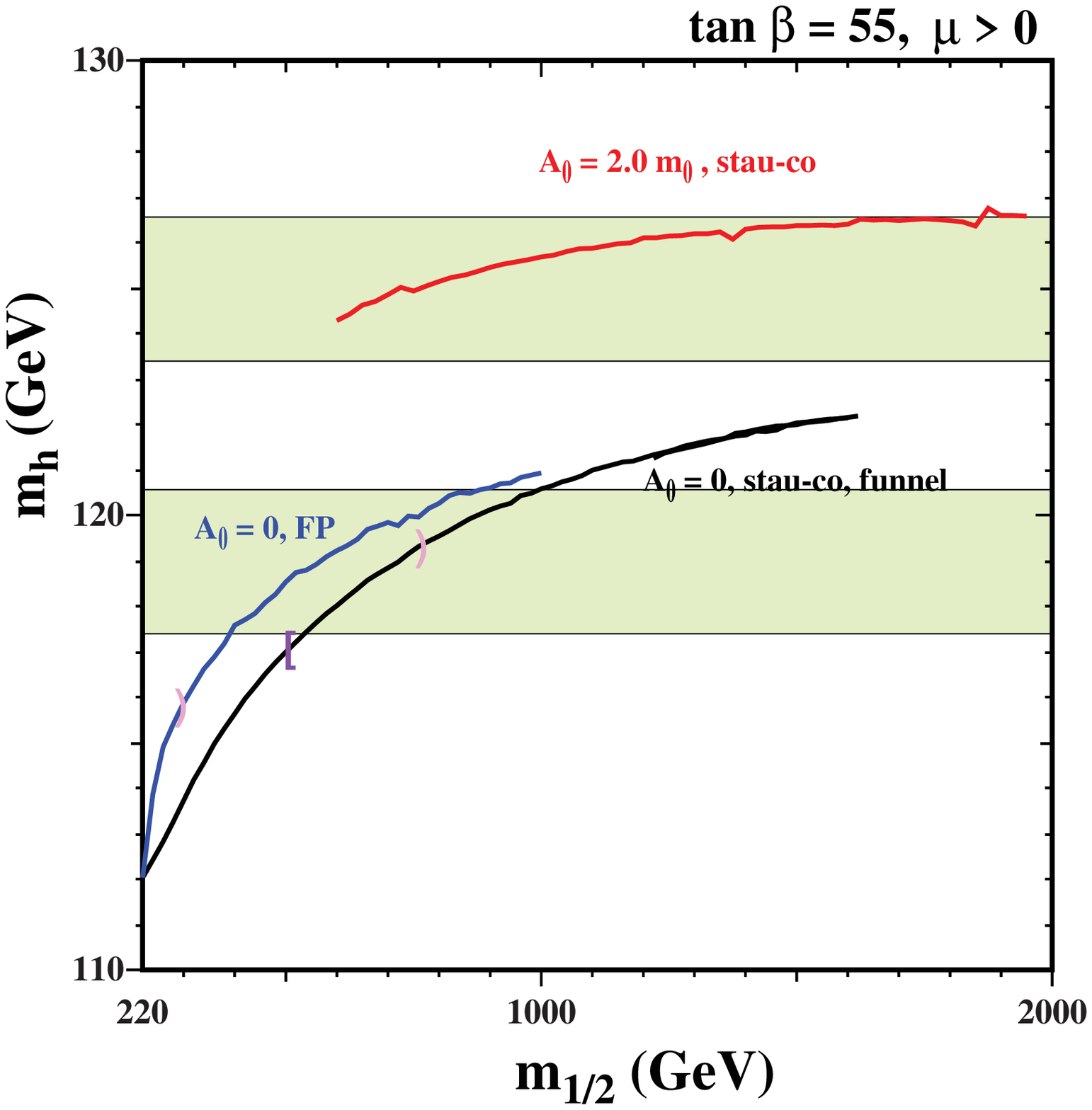,height=3in}
\epsfig{file=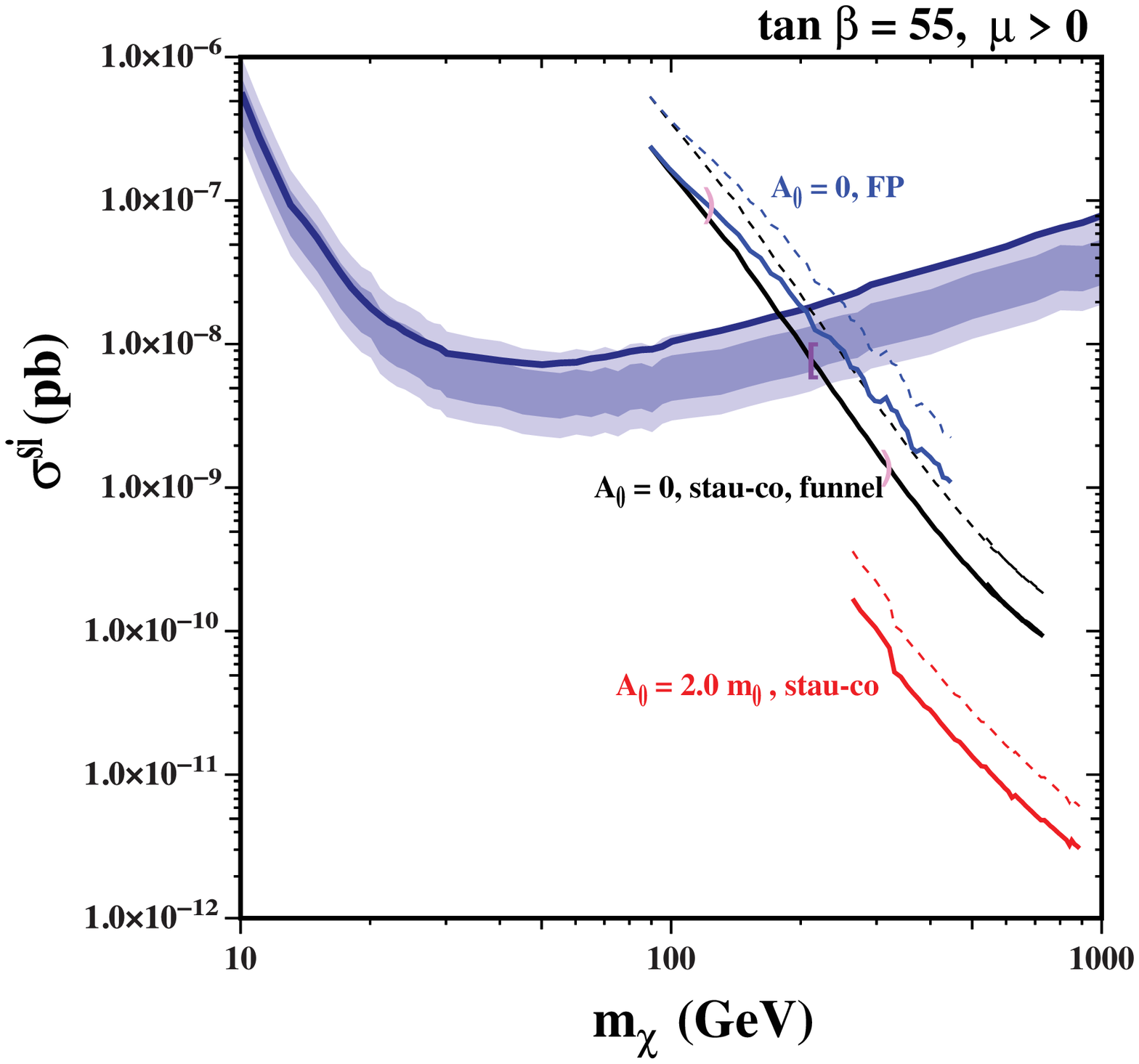,height=3in}
\hfill
\end{minipage}
\vspace{0.1in}
\caption{
{\it
Left panel: $m_h$ as calculated using {\tt FeynHiggs} (showing the bands $m_h = 119 \pm 1.5$~GeV and $125 \pm 1.5$~GeV)
and right panel: spin-independent elastic $\chi-p$ scattering cross section (showing the XENON100 exclusion~\protect\cite{XENON100}
as in Fig.~\protect\ref{fig:10strips}), along WMAP strips for
$\tb = 55$ - the ${\tilde \tau}_1 - \chi$ coannihilation strips for $A_0 = 0$ (\protect\ref{55stau0}) 
(black) and $A_0/m_0 = 2.0$ (\protect\ref{55stau2.0}) (red), and
the focus-point strip for $A_0 = 0$ (\protect\ref{55fp}) (green).
}}
\label{fig:55strips} 
\end{figure}

The left panel of Fig.~\ref{fig:55strips} displays
the values of $m_h$ along various WMAP strips for $\tb = 55$.
As before, the black line is the ${\tilde \tau}_1 - \chi$ coannihilation for $A_0 = 0$
(\ref{55stau0}), and is actually doubled
for $m_{1/2} > 1200$~GeV before terminating at $m_{1/2} \sim 1600$~GeV, corresponding to the two sides of the
rapid-annihilation funnel, though the corresponding values of $m_h$
are very similar. In this case, we see that a range of $m_{1/2}$ is compatible with $g_\mu - 2$
as well as the LHC missing-energy searches and $m_h = 119$~GeV. The red line is for the ${\tilde \tau}_1 - \chi$ 
coannihilation strip with $A_0/m_0 = 2.0$ (\ref{55stau2.0}), as we
do not find generic consistent solutions for $\tb = 55$ and $A_0/m_0 = 2.5$.
As in Fig.~\ref{fig:10strips}, the blue line is for the focus-point strip with $A_0 = 0$ (\ref{55fp}).
We see that $m_h = 119$~GeV is compatible with the $A_0 = 0$ coannihilation
strip for $m_{1/2} > 600$~GeV, and with the $A_0 = 0$ focus-point strip for
$m_{1/2} > 400$~GeV. However, only the ${\tilde \tau}_1 - \chi$ coannihilation strip
with $A_0/m_0 = 2.0$ is compatible with $m_h = 125$~GeV, and this for all values
of $m_{1/2}$.

The right panel of Fig.~\ref{fig:55strips} displays the corresponding elastic scattering
cross section along the same strips. As in Fig.~\ref{fig:10strips}, the focus-point strip
yields the highest cross section, and as there and in Fig.~\ref{fig:40strips}, the cross
section for $A_0 = 0$ is larger than that for $A_0/m_0 > 0$. This time, the cross
section for $A_0/m_0 = 2.0$ is smaller than that for $A_0 = 0$ by more than an
order of magnitude, and is again always $< 10^{-9}$~pb, even for the optimistic
value $\Sigma_{\pi N} = 64$~MeV. Along the focus-point strip, on the other hand,
the cross section could be as large as the XENON100 upper limit.

\section{Summary}

We have discussed in this paper the interplay between a hypothetical measurement
of the mass of the Higgs boson and spin-independent elastic dark matter scattering,
in the context of WMAP strips in the $(m_{1/2}, m_0)$ planes of the CMSSM. In the
past, it has been common to discuss planes with $A_0 = 0$ and various values of
$\tan \beta \in [10, 55]$. However, previous studies~\cite{ENOS,mc7.5,post-mh} have shown
that $A_0 > 0$ may be preferred, so we have explored this possibility in this paper.
Among the examples we consider is a ${\tilde t_1} - \chi$ coannihilation strip, a
possibility that does not arise if $A_0 = 0$, and which has not been extensively
studied in the dark matter detection literature.

Positive values of $A_0$ generally yield larger values of $m_h$ than for $A_0 = 0$,
which may be preferred in light of the LHC `hint' that $m_h \sim 125$~GeV, though
$m_h \sim 119$~GeV may still be a possibility. As could be anticipated from previous
studies, only limited portions of the WMAP strips are compatible with $m_h \sim 125$~GeV,
whereas larger portions are compatible with $m_h \sim 119$~GeV. In addition to
${\tilde \tau_1} - \chi$ coannihilation strips with $\tan \beta \sim 40$ or more and
$A_0 \sim 2 m_0$ or more, which are reflected in Figs.~2 and 3 of~\cite{mc7.5},
we also find that some portion of the ${\tilde \tau_1} - \chi$
coannihilation strip for $\tan \beta = 10$ may also be compatible with $m_h \sim 125$~GeV
within the {\tt FeynHiggs} uncertainty of $\pm 1.5$~GeV if $A_0$ is very large, e.g., $A_0 = 3000$~GeV,
$m_{1/2} \sim 900$~GeV and $m_0 \sim 350$~GeV. Such points would populate the low-$\tb$
tail of the 68\% CL region in the  CMSSM $(\tb, m_{1/2})$ plane shown in Fig.~3 of~\cite{mc7.5}.
On the other hand, most supergravity
models have $A_0 = c. m_0$ with $c \in [-3, 3]$, so this example might not arise in such
scenarios.

CMSSM models lying along WMAP-compatible ${\tilde \tau_1} - \chi$ coannihilation strips 
with $A_0/m_0 > 0$ generally have lower spin-independent elastic dark matter scattering
cross sections than the corresponding cases with $A_0 = 0$. Some models with low $m_\chi$
and $m_h$ are already excluded by the XENON100 upper limit on dark matter scattering, but 
models with $m_h \sim 125$~GeV generally yield cross sections well below this limit, typically
$< 10^{-9}$~pb. It will be interesting to use some of the strips discussed here to benchmark other astrophysical
dark matter strategies, e.g., indirect searches for $\chi - \chi$ annihilations that yield energetic
neutrinos or photons. However, the general (loose) correlation between elastic scattering and
relic annihilation suggests that the rates for such processes may also be suppressed in many
models compatible with $m_h \sim 125$~GeV.

\section*{Acknowledgements}

This work has been supported in part by
the London Centre for Terauniverse Studies (LCTS), using funding from
the European Research Council 
via the Advanced Investigator Grant 267352. 
The work of K.A.O. is also supported in part by DOE grant
DE-FG02-94ER-40823 at the University of Minnesota.

\end{document}